\documentclass[12pt]{article}

\usepackage{newtxmath} 
\usepackage{graphicx}
\usepackage{epsf}
\usepackage{color}
\usepackage{geometry}
\usepackage{booktabs}
\usepackage[american]{babel}
\usepackage{xcolor, soul}
\sethlcolor{yellow}
\usepackage{color}
\usepackage{multirow}
\usepackage{hyperref}
\usepackage{subcaption}
\usepackage{float}

\usepackage{microtype}
\usepackage{comment}

\definecolor{DarkBlue}{rgb}{0.1,0.1,0.5}
\definecolor{Red}{rgb}{0.8,0.1,0.1}
\definecolor{gmugreen}{RGB}{0,112,51}
\definecolor{gmugold}{RGB}{255,204,51}
\definecolor{LightBlue}{RGB}{90,172,240}
\definecolor{LightYellow}{RGB}{234,209,20}
\definecolor{GreenIndigo}{RGB}{229,229,101}
\renewcommand{\epsilon}{\varepsilon}
\renewcommand{\rho}{\varrho}
\renewcommand{\phi}{\varphi}

\newcommand{\N}{{\mathbb N}}

\newcommand{\beanum}{\begin{eqnarray}}
\newcommand{\eeanum}{\end{eqnarray}}
\newcommand{\bea}{\begin{eqnarray}}
\newcommand{\eea}{\end{eqnarray}}
\newcommand{\bsa}{\begin{subeqnarray}}
\newcommand{\esa}{\end{subeqnarray}}
\newcommand{\be}{\begin{equation}}
\newcommand{\ee}{\end{equation}}

\usepackage{wrapfig}
\definecolor{MatlabRed}{rgb}{1,0,0}
\definecolor{MatlabGreen}{rgb}{0,0.5,0}
\definecolor{MatlabBlue}{rgb}{0,0,1}
\definecolor{MatlabMagenta}{rgb}{1,0,1}

\usepackage{ragged2e}

\usepackage{authblk}

\title{A Numerical Comparison of Petri Net and Ordinary Differential Equation SIR Component Models}
\author[1]{Trevor Reckell} 
\author[2]{Bright Kwaku Manu}
\author[3]{Beckett Sterner}
\author[1]{Petar Jevti\'c} 
\author[4]{Reggie Davidrajuh}
\affil[1]{School of Mathematical and Statistical Sciences, Arizona State University, 901 S. Palm Walk, Tempe, AZ 85287-1804, USA}
\affil[2]{School of Computing and Augmented Intelligence, Arizona State University, 699 S Mill Ave, Tempe, AZ 85281}
\affil[3]{School of Life Sciences, Arizona State University, Tempe, USA}
\affil[4]{Faculty of Science and Technology, University of Stavanger, Stavanger, Norway}
\date{\today}

\begin{document}
\maketitle

\begin{abstract} 
Petri Nets are an increasingly used modeling framework for the spread of disease across populations or within an individual. For example, the Susceptible-Infectious-Recovered (SIR) compartment model is foundational for population epidemiological modeling and has been implemented in several prior Petri Net studies. While the SIR model is typically expressed as Ordinary Differential Equations (ODEs), with continuous time and variables, Petri Nets operate as discrete event simulations with deterministic or stochastic timings. We present the first systematic study of the numerical convergence of two distinct Petri Net implementations of the SIRS compartment model relative to the standard ODE. In particular, we introduce a novel deterministic implementation of the SIRS model using dynamic transition weights in the GPenSIM package and stochastic Petri Net models using SPIKE. We show how rescaling and rounding procedures are critical for the numerical convergence of Petri Net SIR models relative to the ODEs, and we achieve a relative root mean squared error of less than 1\% compared to ODE simulations for biologically relevant parameter ranges. Our findings confirm that both stochastic and deterministic discrete time Petri Nets are valid for modeling SIR-type dynamics with appropriate numerical procedures, laying the foundations for larger-scale use of Petri Net models. 

\end{abstract}


\section{Introduction}
\label{intro-sec}

Petri Nets (PNs) are a discrete event mathematical formalism that has become an increasingly popular modeling tool in epidemiology and other areas of the life sciences. Recent applications include models of Covid-19 spread within and across countries \cite{yang2021rational, connolly2022epidemic}, pertussis vaccination \cite{castagno_computational_2020}, social-ecological systems \cite{Pommereau2024}, rumor propagation in social networks \cite{wang_modeling_2015}, and the behavior of network motifs \cite{aduddell_compositional_2024}. Key features of the Petri Net formalism include the ability to construct modular, combinatorial models \cite{herajy_design_2024, aduddell_compositional_2024, libkind2022algebraic};  the ease of visualizing and interpreting Petri Net models for those without a strong background in mathematics; and advances in software implementations that enable large-scale simulation including deterministic and stochastic behaviors that account for real-world aspects and phenomenon \cite{davidrajuh2018-book, davidrajuh2021petri, davidrajuh2023colored, chodak2021spike}. 

However, some essential questions remain about how Petri Net implementations of Susceptible-Infectious-Recovered (SIR) models relate to their standard formulations using Ordinary Differential Equations (ODEs) in epidemiology. While prior theoretical results indicate that Petri Net SIR models can be made to converge asymptotically to the behavior of ODE and Stochastic Differential Equation (SDE) counterparts \cite{beccuti_analysis_2014}, there has been little systematic investigation of the numerical conditions under which this occurs using contemporary software tools. For example, the fitting and forecasting abilities of Petri Nets compared to ODEs are not considered in \cite{koch2010modeling}, \cite{peng2021modeling}, \cite{chen2003quantitative}, and \cite{libkind2022algebraic}. Understanding precisely when and how Petri Net SIR models may diverge from ODE or SDE systems is critical for large-scale combinatorial modeling efforts \cite{herajy_design_2024,aduddell_compositional_2024,libkind2022algebraic,chiaradonna_mpat_2024}. 

In this work, we investigate the numerical convergence of two Petri Net implementations to an ODE SIRS model. The second S in SIRS indicates that individuals may become susceptible again after recovering from an infection. We present a novel implementation of SIR-type models using dynamic arc weights for transitions in a Petri Net model with deterministic time steps using the GPenSIM toolkit. We also analyze a Continuous-Time Markov Chain (CTMC) Petri Net with stochastic time steps using the Spike toolkit\footnote{\url{https://github.com/trevorreckell/Numerical-Comparison-of-PN-vs-ODE-for-SIR}}. We evaluate the behaviors of these Petri Net models to numerical solutions of ODE model using a standard MatLab differential equation solver.

We find that both Petri Net models can be made to converge to within $1\%$ relative root mean squared error (RRMSE) of the ODE using appropriate rescaling of population size in the case of the CTMC Petri Model and a combination of population size and time step in the dynamic arc weight model. We also find that the choice of numerical rounding procedures has a significant impact on the RRMSE of the dynamic arc weight model when the infected population is close to zero. Our results, therefore, provide an invaluable guide for the numerical simulation of SIR dynamics under biologically realistic parameter values.

In section \ref{sec:fund PN}  we introduce some basics of the Petri Net formalism and introduce two different ways that Petri Nets can be simulated. In section \ref{sec:PN imp} and \ref{sec:models} we discuss specific ways in how PN simulations can be improved in relation to a known system of ODEs within the software framework that we have laid out. Lastly, in section \ref{sec:results} the PN simulation methods are compared to ODEs and results of how the PN simulation improvement methods affect the computation time are shown. In section \ref{sec:conlusion}  we conclude how the methods we laid out set the groundwork for proper implementation of Petri Nets.

\section{Methodology}
\label{methods-sec}
Our approach is based on the numerical comparison of PN models to ODEs, which are a standard modeling framework used in fields such as epidemiology, drug pharmacokinetics, and cancer biology \cite{auger2008structured}, \cite{lu2021neural}, \cite{reckell2021modeling}, and \cite{aguda2008microrna}. The SIR model is the most commonly used in epidemiology, and it also serves as the foundation for a large family of more complex models. Kermack and McKendrick initially derived the model in 1927 \cite{kermack1927contribution} by structuring the total host population into into three different "compartments": $S$ denoting the susceptible population, $I$ denoting the infected population, and $R$ denoting the recovered population. Equations \ref{eq:S_SIRS}-\ref{eq:R_SIRS} describe the resulting system behavior, with the parameter $\beta$ representing the rate at which susceptible populations become infected and $\gamma$ representing the rate at which the infected population recovers per unit of time.

The classical SIR model assumes that there is no birth, death, immigration, or emigration, that populations mix homogeneously, that new infections are not dependent on other factors besides $\beta SI$, and that there is an exponential waiting time for events to happen in each compartment. In practice today, the original SIR model is typically used as a basis for constructing more complex models that include disease-specific dynamics and various ways of treating or preventing the disease. 

For example, the SIR model can easily be adapted to the SIRS model by adding the term $\delta R$, which represents the rate at which hosts become re-susceptible per unit of time. By adding $\delta R$ to the $S$ compartment and subtracting it from the recovered population $R$ compartment, we get Equations \ref{eq:S_SIRS}-\ref{eq:R_SIRS}. Note that when $\delta=0$, Equations \ref{eq:S_SIRS}-\ref{eq:R_SIRS} are equivalent to an SIR model.
\begin{eqnarray}
\frac{dS}{dt} &=& \delta R-\beta SI\label{eq:S_SIRS} \\
\frac{dI}{dt} &=& \beta S I- \gamma I \label{eq:I_SIRS} \\
\frac{dR}{dt} &=& \gamma I - \delta R.\label{eq:R_SIRS} 
\end{eqnarray}

\subsection{Fundamentals of Petri Nets} \label{sec:fund PN}
A Petri Net graph, or Petri Net structure, is a weighted bipartite graph \cite{cassandras2008introduction} defined as $n$-tuple 
    $(P, T,A,w,x)$ where, 
\begin{itemize}
    \item $P$ is the finite set of places (one type of vertex in the graph).\vspace{-3mm}
    \item $T$ is the finite set of transitions (the other type of vertex in the graph). \vspace{-3mm}
    \item $A$ is the set of arcs (edges) from places to transitions and from transitions to places in the graph $A \subset (P \times T) \cup (T \times P)$. \vspace{-3mm}
    \item $M_0$ is the initial state (also known as marking), $M_0 = [m_1, m_2, \ldots, m_n]$, where $m_i$ is the number of tokens in place $p_i$ .\vspace{-3mm}
    \item $w$ : A → \{1, 2, 3, . . .\} is the weight function on the arcs.\vspace{-3mm}
    \item $x$ is a marking of the set of places P; $x = [x(p_1), x(p_2), . . . , x(p_n)$] $\in N^n$ is the row vector associated with $x$.
\end{itemize}
Tokens are assigned to places, with the initial assignment being the initial marking. The number of tokens assigned to a place is an arbitrary non-negative integer but does not necessarily have an upper bound. A transition $t_j \in T$ in a Petri Net is said to be enabled if $x(p_i) \geq w(p_i, t_j)$ for all $p_i \in I(t_j)$, where $I(t_j)$ is the set of input arcs from places to $t_j$. This allows us to define the state transition function, $f : \N^n \times T \to \N^n$, such that transition $t_j \in T$ fires if and only if $x(p_i) \geq w(p_i, t_j)$ for all $p_i \in I(t_j)$. If $f(x, t_j)$ is defined, then we set $x' = f(x, t_j )$, where $x'(p_i) = x(p_i) - w(p_i, t_j) + w(t_j, p_i), i= 1, . . . , n$. In simple terms, a transition is enabled if the number of tokens in all places connected to that transition via an incoming arc is greater than or equal to the arc weight for the respective arc connected to the transition. 

In order to be precise about the relative time scales of firing events in PN models versus ODE models, we define $\tau$ to be the number of steps per unit of time in a Petri Net with deterministic time steps and firings. The graph then can be defined as $(P,T,A,w,x,\tau)$ as applied to the deterministic implementation of a Petri Net used in GPenSIM \cite{davidrajuh2018-book,davidrajuh2018optimizing}, where $\frac{1}{\tau}$ is called sampling frequency and Davidrajuh defines $\tau$ as the continuity of the sequence of execution.

\begin{figure}[H]
    \centering
    \includegraphics[width=0.38\textwidth]{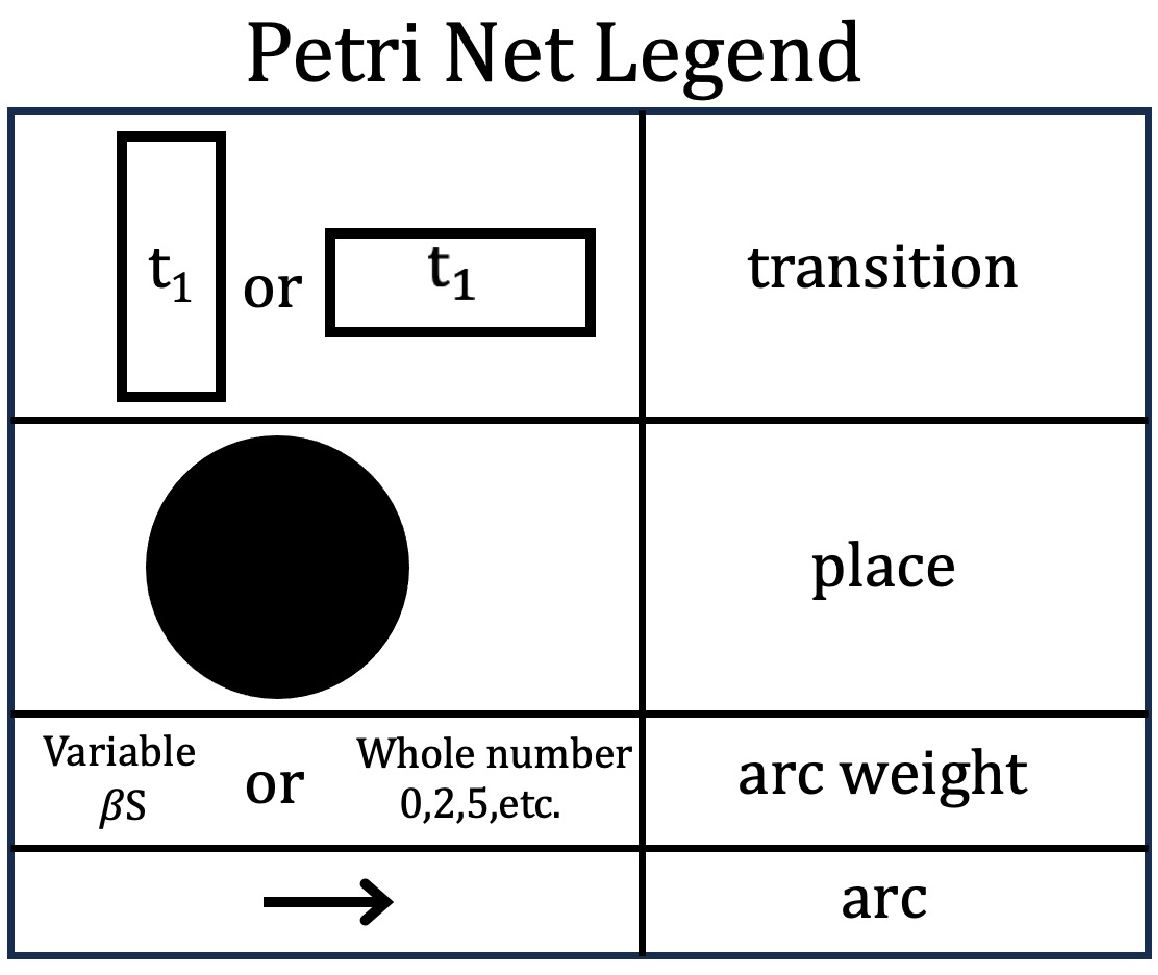} 
  \caption{Petri Nets model formalism elements}\label{fig:P_Legend}
\end{figure}

The last fundamental element of Petri Nets for our purposes is how the diagrams are drawn. We use the visual elements shown in Figure \ref{fig:P_Legend} to describe the logic of both deterministic Petri Nets with variable arc weights and stochastically firing Petri Nets with fixed arc weights.

\subsection{Petri Net Implementations} \label{sec:PN imp}

A variety of Petri Net simulation and software tools are freely available online, all with relative advantages and disadvantages \cite{kumbhar2023review,thong2015survey}. The most commonly used tools, such as GPenSIM, Snoopy, CPNTools, and Spike, can handle a range of different Petri Net types. We will focus on two tools, GPenSIM and Spike, based on their robust features for scalable modeling and simulation analysis and command-line interfaces for scripting. 

All of our code used in this paper can be found at \url{https://github.com/trevorreckell/Numerical-Comparison-of-PN-vs-ODE-for-SIR}. From here, there are sections in the main simulation file of \texttt{simple\_pn\_SIRv1\_preprint.m} to reproduce each figure and alter the code for individual purposes.

\subsubsection{GPenSIM}
The General-purpose Petri Net Simulator (GPenSIM) package is a new tool for the MATLAB platform. GPenSIM is designed for modeling, simulation, and performance analysis of discrete systems. There were many reasons for developing GPenSIM \cite{davidrajuh2018-book} with the primary motivation being to create a system that is easy to learn and use for both industrial and academic applications. Interoperability was also a priority so that GPenSIM can model discrete systems in different domains such as Production and Mechanical Engineering, Industrial Engineering, Computer Science and Engineering, etc., taking full advantage of MATLAB's numerous functions in diverse toolboxes. Another goal was to provide a simple core engine with an extensible interface so that users can modify GPenSIM functions or create new functions to model their systems.

The initial version of GPenSIM was developed for simple analyses of place/transition Petri Nets \cite{davidrajuh2018-book}. Later, coloring capability (Colored Petri Nets) was added so that real-life systems could be modeled \cite{davidrajuh2023colored}. For modeling large real-life systems, modular modeling capability was added \cite{davidrajuh2021petri}. 

Petri Nets implemented by GPenSIM are fundamentally deterministic, and GPenSIM allows Petri Nets with non-dynamic arc weights, which we call static Petri Nets. These static Petri Nets are defined as to have arcs that are constant throughout the entire simulation. At the start of the simulation, a static Petri Net graph must be defined in a separate file (Petri Net Definition File (PDF)), and this file will be used throughout the simulation. The limitations of a static Petri Net can be overcome using a dynamic arc weight Petri Net implemented using an iterative approach as follows. One starts with a static Petri Net graph with some initial arc weights (initial definition file). The PN simulation is run for one time step, and the resulting places markings are then used to calculate updated arc weights, which are fed into a new PDF file for the next iteration. Since the PN definition file must be recreated for each iteration step, execution time may be an important issue to consider.

\subsubsection{Spike}
Spike is a Petri Net simulation framework designed to model dynamic systems using stochastic and deterministic approaches. It emphasizes reproducibility for stochastic simulations by using repeatable configuration files, denoted by the ``.spc" file type \cite{chodak2021spike}. Spike also supports diverse simulation methods for Continuous-Time Markov Chain (CTMC) processes, including Gillespie’s direct method for exact stochastic modeling \cite{gillespie1977exact}, tau-leaping and delta-leaping for efficient approximations \cite{rohr2018discrete}, and deterministic ODE solvers for large-scale systems. By definition, a CTMC describes a discrete set of system states where the future state of a system depends only on its current state (memoryless), and the time between transitions to another state follows an exponential distribution. 

Spike also has multiple notable features and capabilities\cite{chodak2021spike}. Efficient simulation is one important feature, since Spike supports the automated execution of large sets of simulations, which can be run sequentially or in parallel. Spike also supports the simulation of stochastic, continuous, and hybrid Petri Nets, accommodates colored and uncolored variants, and leverages its integration with the broader PetriNuts software framework. Spike is also designed to support various use cases, including benchmarking, adaptive model simulations, parameter scans, and model optimization. However, the underlying code of Spike is not publicly available, which limits one ability to inspect and debug the exact steps being executed by different functions. 

\subsection{Petri Net SIRS Models in Spike and GPenSIM} \label{sec:models}

\begin{figure}[H]
    \centering
    \includegraphics[width=0.7\textwidth]{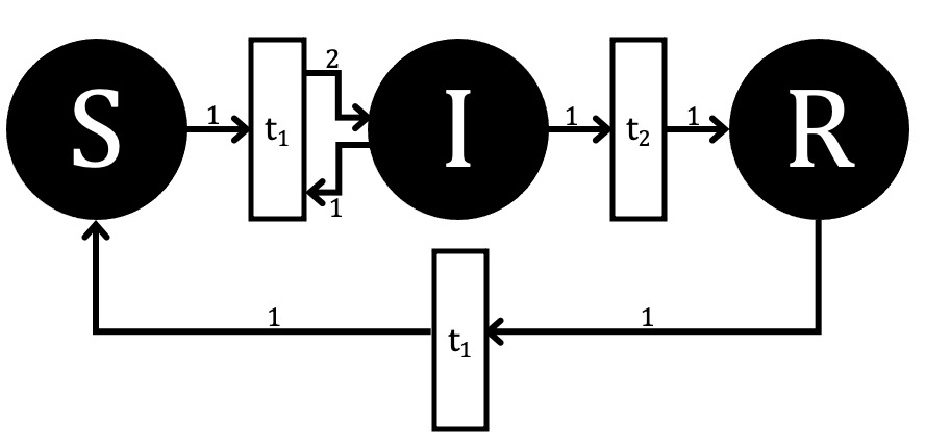} 
  \caption{Continuous-time, Markov Chain SIRS model with stochastic firing times and fixed arc weights, as implemented in Spike}
         \label{fig:SIR_v2}
\end{figure}

\begin{figure}[H]
    \centering
    \includegraphics[width=0.7\textwidth]{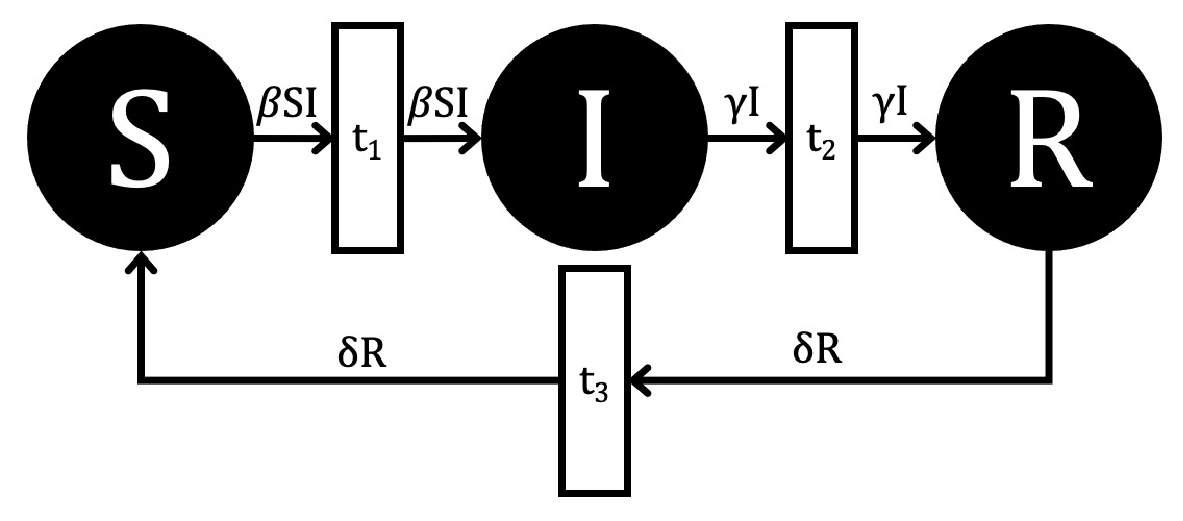}
    \caption{Discrete-time, deterministic SIRS model using fixed time steps and dynamic arc weights, as implemented in GPenSIM.}
    \label{fig:SIRS_PN}
\end{figure}

We now turn to describe the SIRS model in both a deterministic, variable arc weight form using GPenSIM and a stochastic, fixed arc weight form using Spike. In both cases, a PN model can be stated that preserves the corresponding ODE model's assumptions (such as waiting times and closed population) and biological dynamics (such as positive populations and susceptible populations always able to be infected). However, the implemented PN models have different theoretical arcs, arc weights, and resulting dynamics. Figures \ref{fig:SIR_v2} and \ref{fig:SIRS_PN} illustrate these differences relative to the SIRS ODE Equations \ref{eq:S_SIRS}-\ref{eq:R_SIRS}.  

The layout in Figure \ref{fig:SIR_v2} reflects the application of the mass action equation from chemical reactions to determine the pattern of arcs and weights that correspond to the ODE system \cite{segovia2022petri,baez2012quantum}. Although the arc weights are shown as static, the frequency at which the arcs fire is dynamic as a function of $\beta S I$ over time because the Petri Net is implemented as a CTMC. The arc weights themselves in this layout therefore describe only the stoichiometry of tokens from each place involved in a transition. 

Shifting to a deterministic PN, we have fixed time steps and the arc weights change dynamically with each time step. A key change from the CTMC layout in Figure \ref{fig:SIR_v2} is the removal of the arc from place $I$ to transition $t_1$. Additionally, the arc weights are now a function of time since their value at a time point depends on the number of tokens in one or more compartments, and they directly represent the flow of tokens between compartments instead of the stoichiometry of reactions. This shift in semantics is matched by having the deterministic PN fire on uniformly separated, discrete time points instead of stochastically in continuous time.

\subsection{Numerical Features of Deterministic PN Model}

We describe three numerical features that need to be addressed for the novel deterministic implementation of the SIRS model using GPenSIM: handling of discrete population sizes, rounding of non-integer arc weights, and choice of simulation time step $\tau$.  

\subsubsection{Discrete Population Sizes} \label{sec:Deal_ex}

Since Petri Net implementations in GPenSIM use discrete token values for the SIR compartments, this may disable the firing of transitions even when the arc weights and place markings are both non-zero, leading the Petri Net dynamics to diverge from the corresponding ODE solution. For example, the arc going into transition 1 has a variable arc weight of $\beta SI$ (see Figure \ref{fig:SIRS_PN}). If $S=1$ and $I>\frac{1}{\beta}$, this arc will not fire even though the one susceptible individual should in principle become infected. We consider two solutions for this biologically implausible scenario.
\begin{figure}[H]
     \begin{center}
  \begin{subfigure}[h]{0.49\textwidth}
         \centering
         \includegraphics[width=1\textwidth]{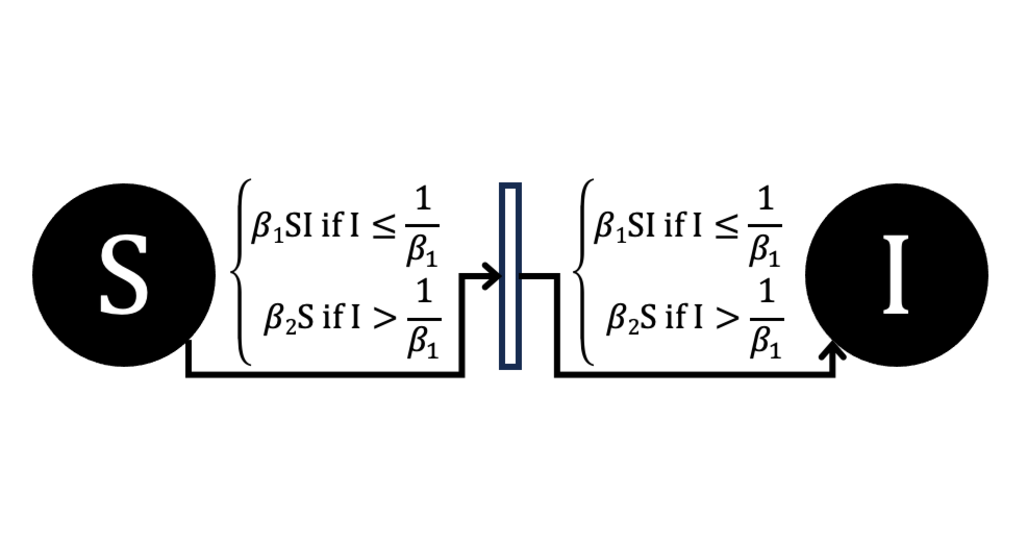}
         \caption{}
         \label{fig:PLow_v1}
     \end{subfigure}
        \hfill
     \begin{subfigure}[h]{0.49\textwidth}
         \centering
         \includegraphics[width=1\linewidth]{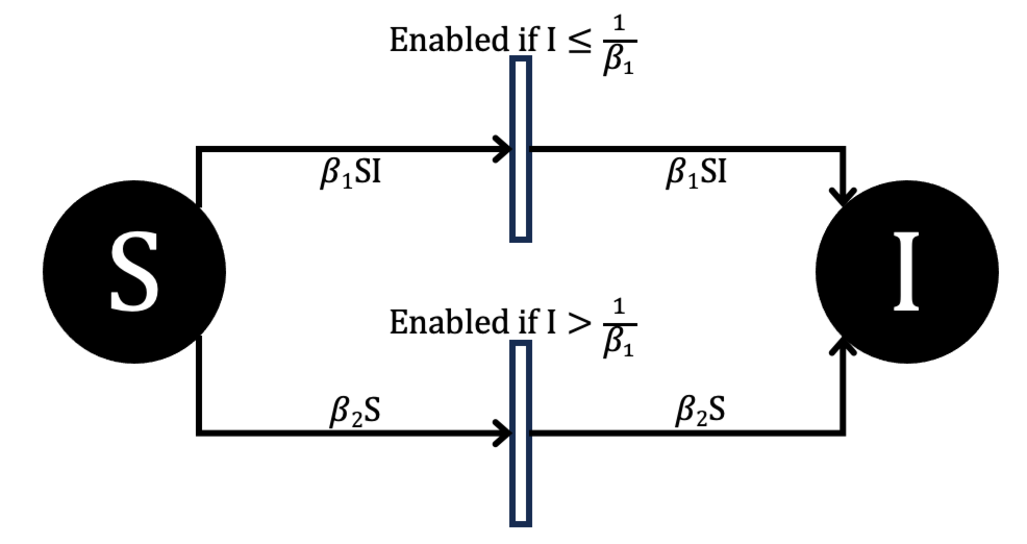}
         \caption{}
         \label{fig:PLow_v2}
     \end{subfigure}
  \end{center}
    \caption{Two options for addressing arcs that become disabled in the deterministic PN model in biologically implausible ways. In Figure  \ref{fig:PLow_v1}, this can be achieved with a simple ``if else" logic statement. In \ref{fig:PLow_v2} an additional transition needs to be installed.}
    \label{fig:Ex_low_lay_optns}
\end{figure}
For example, using the options outlined in Figure \ref{fig:Ex_low_lay_optns}, we can allow for the transition to fire even if the arc weight is initially higher than the place has tokens, specifically if the system is calling for a number of tokens to be moved from place $S$ (susceptible population) to place $I$ (infected population), which is larger than the number of tokens in place $S$. The transition would normally not fire until enough tokens have built up in place $S$ (susceptible population) or the arc weight has changed to a lower value. However, this does not follow the biological dynamics described in the reference ODE model. First, we define $\beta, \beta_1,$ and $ \beta_2$ to have a range of $[0,1]$, as they represent the proportion of individuals who become sick through contact with an infected individual. Thus, we can adapt either \ref{fig:PLow_v1} or \ref{fig:PLow_v2} implementations to allow a firing to occur still, since in both implementations, $\beta_2 S$ is always greater than $S$, since $\beta_2$ is defined as less than one. In figure \ref{fig:PLow_v1}, we allow the formula for the arc weight to switch when we transition $t_1$ is disabled, where this transition being disabled corresponds biologically to not being able to infect the susceptible population. The formula goes from $\beta_1 S I$ to $\beta_2 S$ where $\beta_2=\frac{1}{I+1}$. If $\beta_2\leq\frac{1}{I}$, then transition $t_1$ will be enabled, but the formula for $\beta_2$ could be set to anything in the range of $[0,\frac{1}{I}]$. The formula will depend on the system being modeled and the desired dynamics. The method laid out in figure \ref{fig:PLow_v2} is similar, just with a different transition being enabled if $I>\frac{1}{\beta_1}$, rather than the formula for the arc weight changing for the same transition. Another method of dealing with these extreme values is outlined in the PN time steps per unit time in Section \ref{sec:time_step_meth}. Additionally, population scaling can mitigate the effect this issue has on overall results as seen in section \ref{sec:pop_scal}.

\subsubsection{Integer Arc Weights}
Petri Net simulation software, as a whole, does not have a standardized method for rounding dynamic arc weights, also called dynamic arc weights or arc expressions in some software documentation \cite{melberg2009dynamic, liu2012colored}. Using the standard PN approach, rounding the arc weight is necessary to have the arc weights be positive integers. If positive integer arc weights is difficult or unreasonable for the Petri Net application, one could use continuous weights and token values. Unfortunately, the software options for building scalable, complex models with continuous value arc weights are minimal. Among dynamic arc weight PN software with discrete arc weight values, there is no standardized rounding system for the dynamic arc weights. With this in mind, the question arises about what rounding system should be used. The initial functions we will explore are the ceiling function (weights are rounded up to the nearest positive integer), floor function (weights are rounded down to the nearest positive integer), and standard rounding (weights are rounded down to the nearest positive integer when the tenths place is less than five and up to the nearest positive integer when the tenths place is greater than or equal to five). 

In addition, we propose another rounding method that utilizes standard rounding with the addition of carrying the residual of the rounding process to the next time step. In the next time step, the residual is added to the same arc's weight before that weight is rounded again. This process is repeated each time the dynamic arc weight is recalculated. This process could be done with other rounding functions, like ceiling or floor functions, and we will compare this ``standard + residual" method with many other possible rounding methods more rigorously in future work. These rounding methods will each likely have their own drawbacks in terms of computation time and memory, but our primary goal here is for the resulting PN to most closely mimic that the reference ODE. 

\subsubsection{Time Scales} \label{sec:time_step_meth}

When changing $\tau$ in the deterministic PN, the other parameters implemented are dependent on a respective time unit, so they need to be scaled to ensure they are applied appropriately. For instance, if the susceptible population becomes infected at a rate of 5\% per unit of time, $\beta=0.05$, and we are changing the PN time step to twenty time steps per one unit of ODE time, then we need to scale $\beta$ by $\frac{1}{20}$ giving $\beta=0.05\cdot\frac{1}{20}=0.0025$. With the lower parameter value, this method also has the benefit of avoiding the situation of not allowing the transition to fire as frequently as laid out in Section \ref{sec:Deal_ex}.

With either method, varying time steps or firing times, there will be a significant trade off in computation time. As such, we will calculate the mean computation time when running the model for various time steps to gauge what time step level is necessary given the desired dynamic level and computational power available.

\subsection{Comparison of Petri Net to ODE Models}

We will compare the behaviors of both PN models to the ODE system under the full permitted range of the rate parameters $\beta, \gamma, $and $\delta$. For simplicity, our code \footnote{\url{https://github.com/trevorreckell/Numerical-Comparison-of-PN-vs-ODE-for-SIR}} follows the notation for root mean square error used in MATLAB. We denote the observed data vector as $A$, where $A_i$ is a single vector entry indexed by the time point $i$. The forecasted data is denoted as $F$, where $F_i$ is a single forecasted vector entry indexed by the time point $i$. Finally, $n$ represents the total number of time points being compared. The RRMSE formula is:
\begin{equation*}
    RRMSE=100 \cdot \sqrt{\frac{\frac{1}{n}\sum_{i=1}^n|A_i-F_i|^2}{\sum_{i=1}^n|A_i|^2}} 
\end{equation*}\\
where $A_i$ is the ODE model data, and $F_i$ is the forecasted PN model data in our application. 

We aim to understand how the RRMSE is affected by rescaling key model parameters, specifically the total population size $N$ in both models and the base time step $\tau$ in the deterministic model specifically. These parameters are known to control the relative asymptotic convergence of PN, ODE, and SDE systems for the SIR model \cite{beccuti_analysis_2014}. Since the deterministic, dynamic arc weight implementation of the SIRS model is novel in the PN literature, we give more attention to the details of its implementation than the Spike CTMC model. In particular, we will discuss different rounding strategies for the variable arc weights.

\section{Results} \label{sec:results}

Before comparing the deterministic PN model to the reference ODE system, it is necessary to first address which rounding method is best, since this will affect how the model behaves under rescaling of total population size $N$ and time step $\tau$. 

 The code for all of the models can be found in the GitHub \footnote{\url{https://github.com/trevorreckell/Numerical-Comparison-of-PN-vs-ODE-for-SIR}}. The code contains the model laid out as PN, and the corresponding ODE is defined as a function.  

\subsection{Arc Weight Rounding in GPenSIM Model}
We tested the various rounding methods for parameters $\beta$, $\gamma$, and $\delta$, rate of infection, rate of recovery, and rate of re-susceptibility respectively, as laid out in the SIRS model for ODE and PN at extreme values, including (0,0,0) and (1,1,1), as well as intermediate, biologically plausible values.
\begin{figure}[H]
     \begin{center}
     \textbf{Rounding Error Method Comparison}\par 
     \vspace{3mm}
     \justifying
     \text{\hspace{2mm} $\beta=1$, $\gamma=1$, $\delta=1$               \hspace{28mm} $\beta=0.05$, $\gamma=0.05$, $\delta=0.05$\\ 
     }
     \centering
     \begin{subfigure}[b]{0.49\textwidth}
         \centering
         \includegraphics[width=1\linewidth]{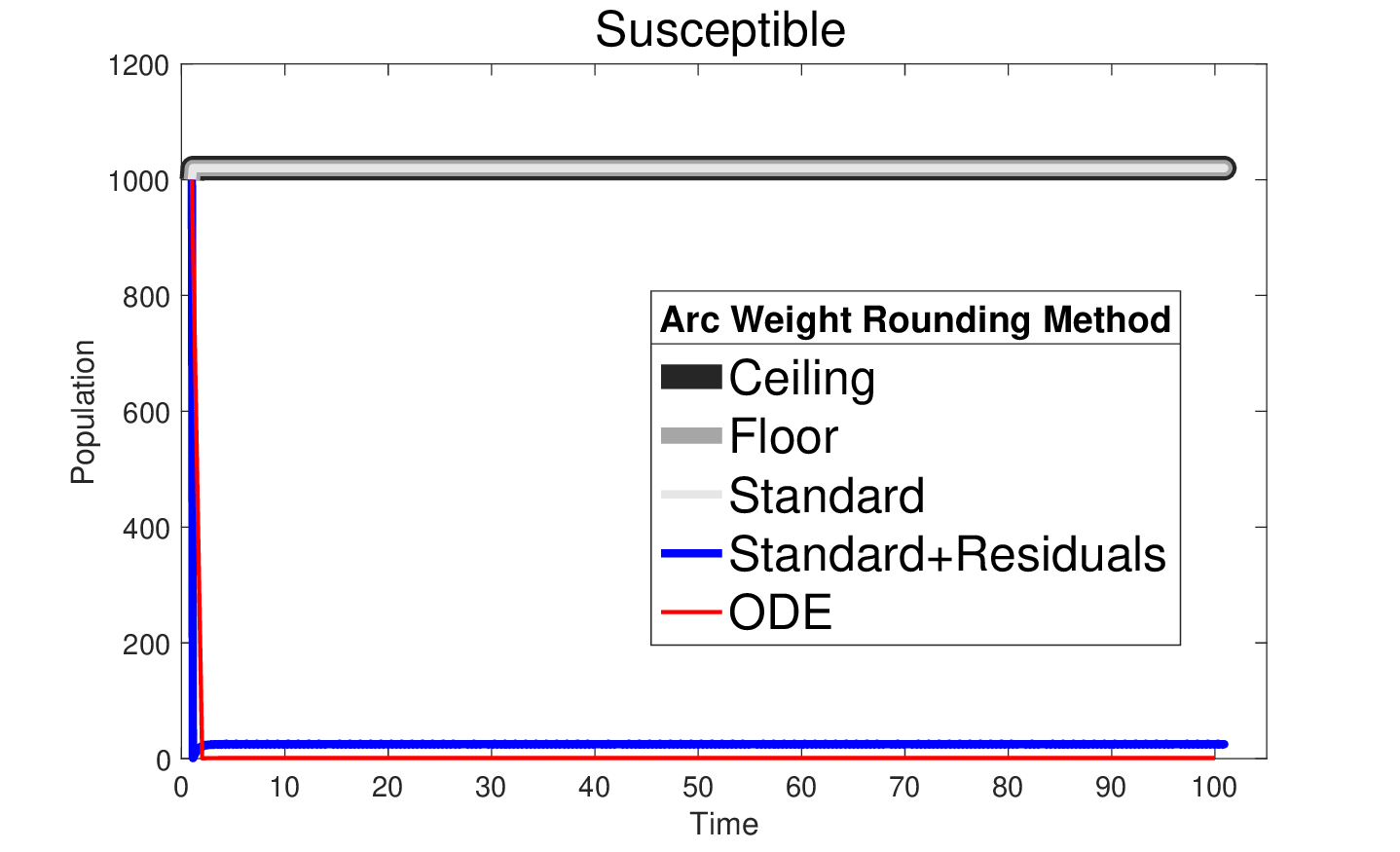}
         \vspace{-8mm}
         \caption{}
         \label{fig:Rounding_1_Sus}
     \end{subfigure}
     \vspace{2mm}
        \hfill  
    \begin{subfigure}[b]{0.49\textwidth}
         \centering
         \includegraphics[width=1\linewidth]{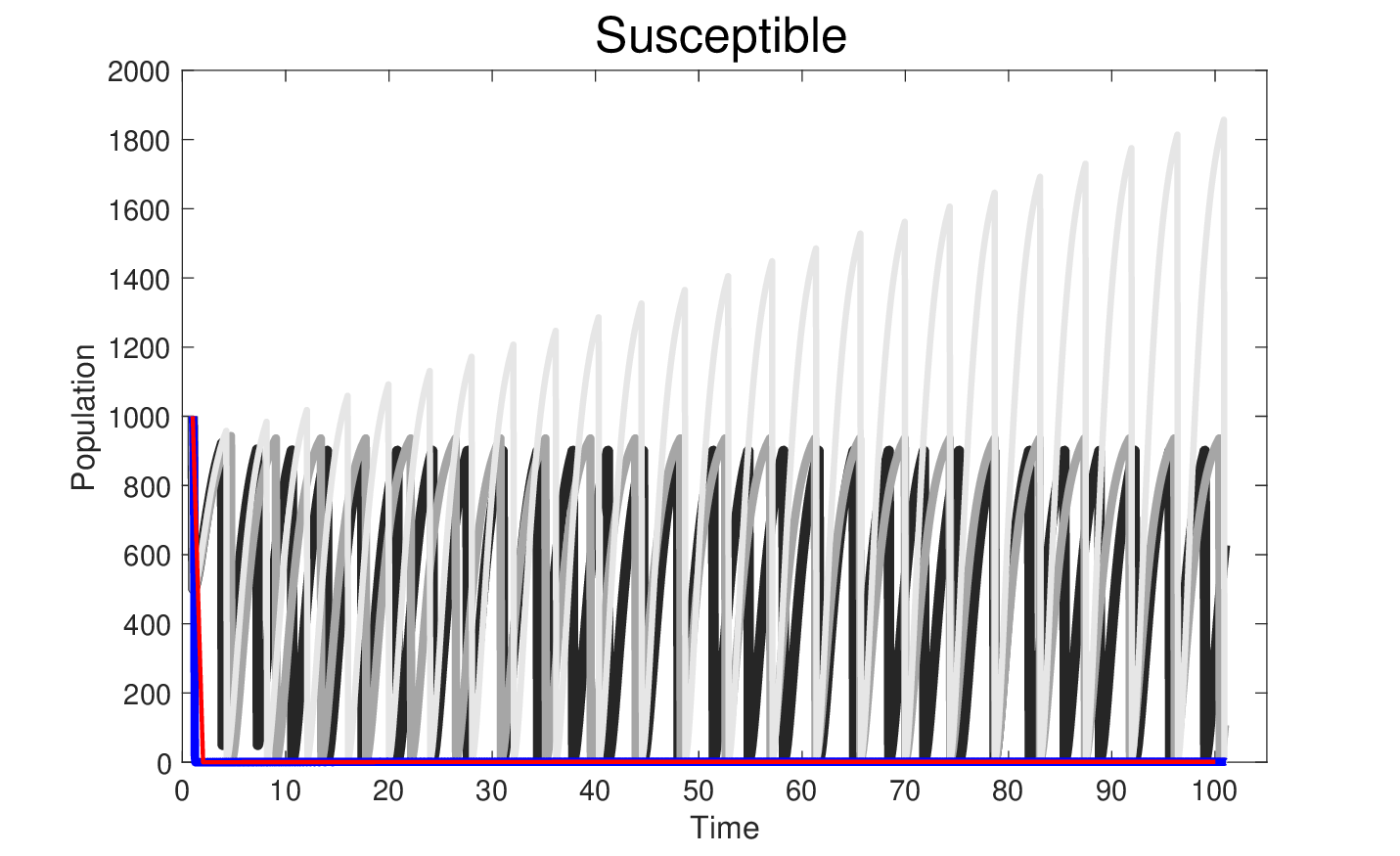}
         \vspace{-8mm}
         \caption{}
         \label{fig:Rounding_05_Sus}
     \end{subfigure}
     \vspace{2mm}
        \hfill
    \begin{subfigure}[b]{0.49\textwidth}
         \centering
         \includegraphics[width=1\linewidth]{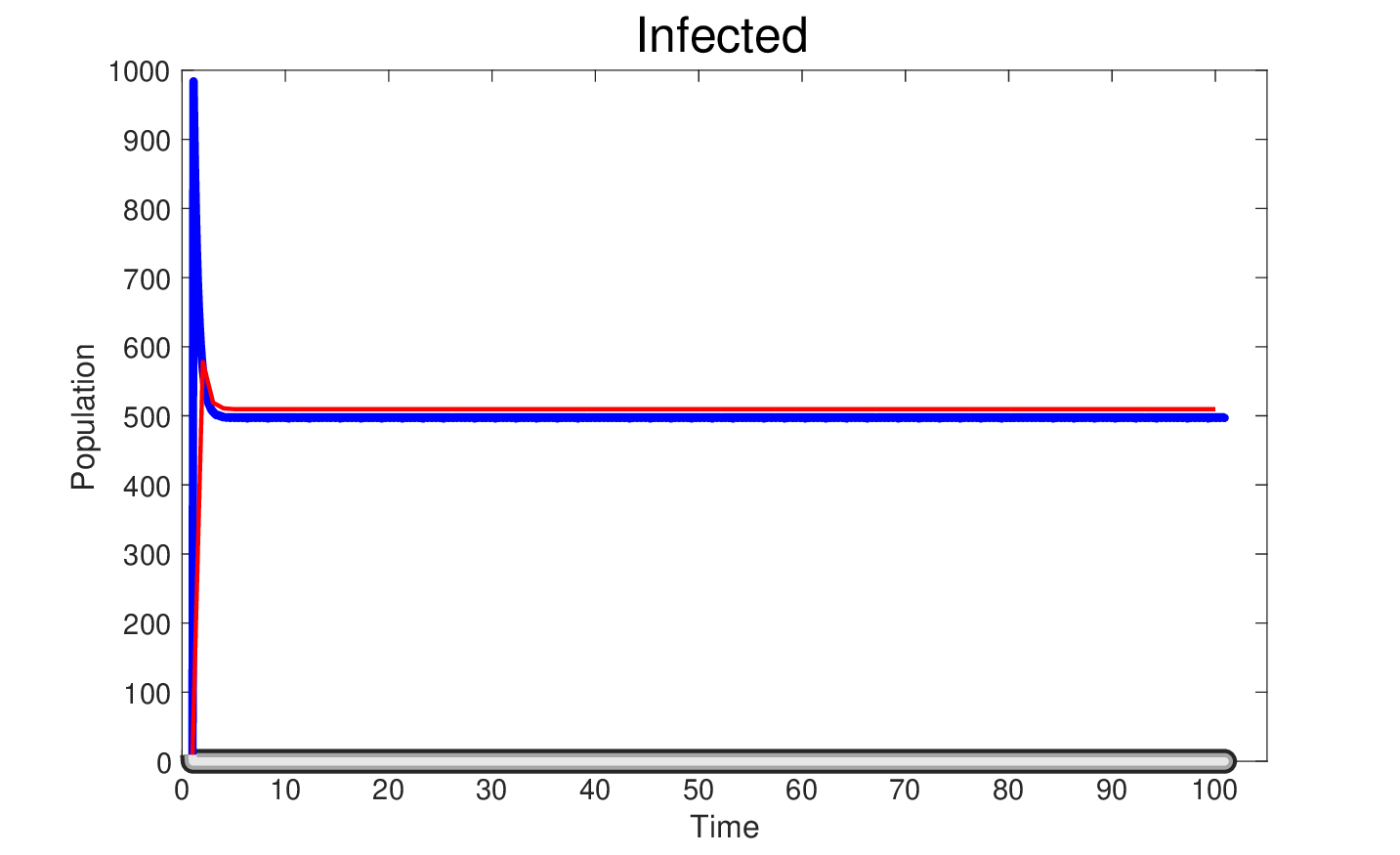}
         \vspace{-8mm}
         \caption{}
         \label{fig:Rounding_1_Inf}
     \end{subfigure}
     \vspace{2mm}
        \hfill
    \begin{subfigure}[b]{0.49\textwidth}
         \centering
         \includegraphics[width=1\linewidth]{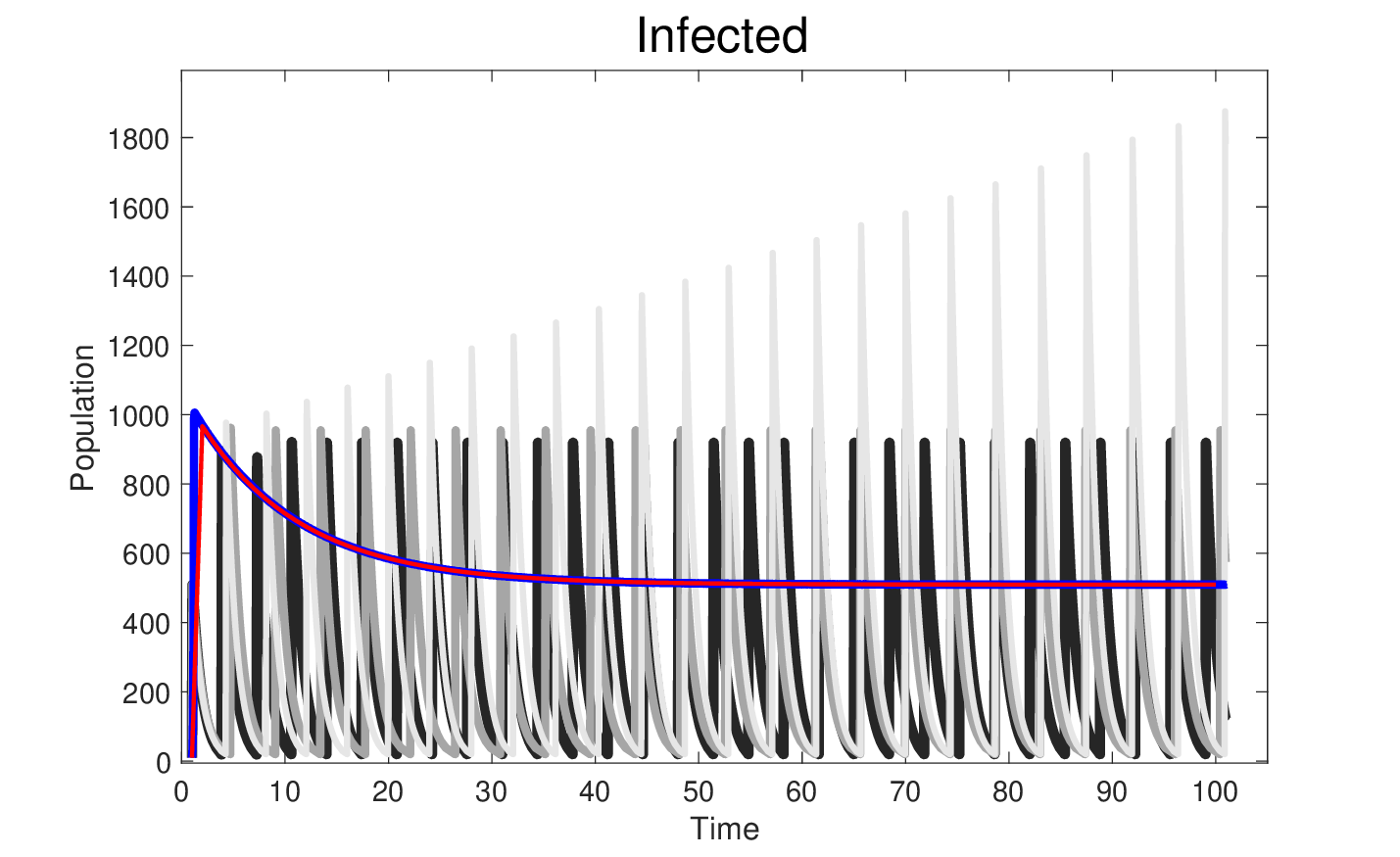}
         \vspace{-8mm}
         \caption{}
         \label{fig:Rounding_05_Inf}
     \end{subfigure}
     \vspace{2mm}
        \hfill
    \begin{subfigure}[b]{0.49\textwidth}
         \centering
         \includegraphics[width=1\linewidth]{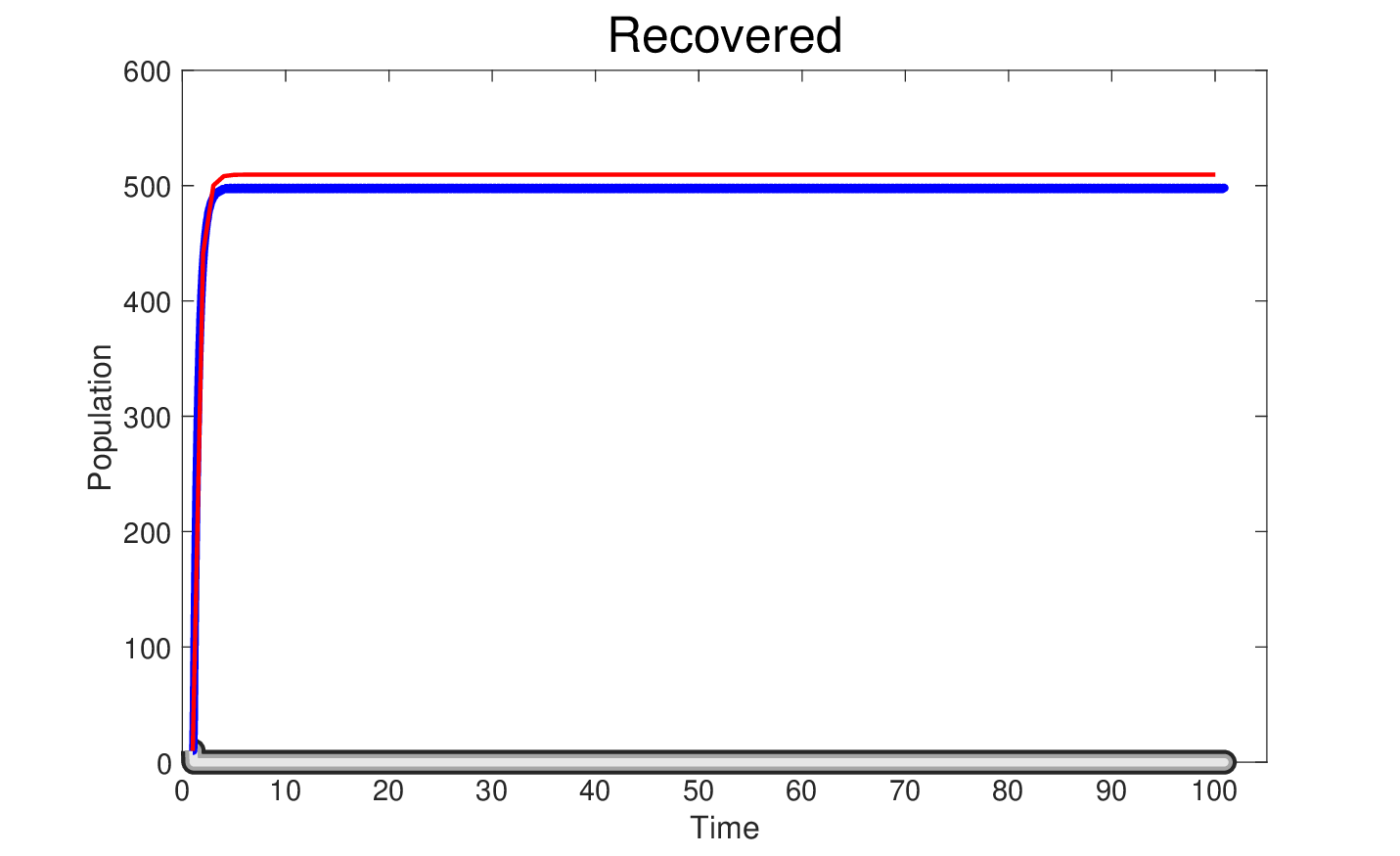}
         \vspace{-8mm}
         \caption{}
         \label{fig:Rounding_1_Rec}
     \end{subfigure}
        \hfill
     \begin{subfigure}[b]{0.49\textwidth}
         \centering
         \includegraphics[width=1\linewidth]{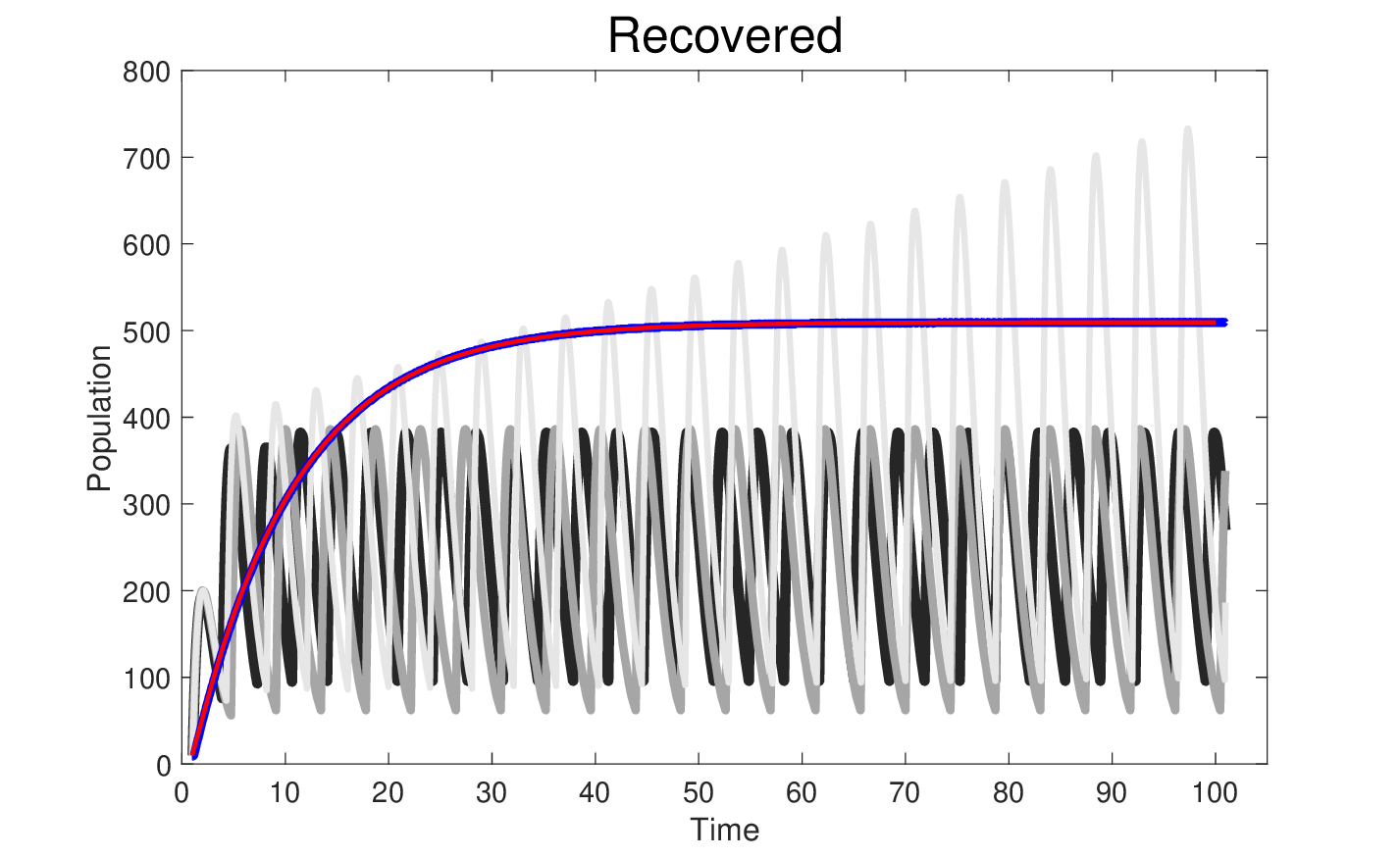}
          \vspace{-8mm}
         \caption{}
         \label{fig:Rounding_05_Rec}
     \end{subfigure}

     \end{center}
     \vspace{-8mm}
 \caption{Comparison of rounding method performance on simulated data. Left column has $\beta=1$, $\gamma=1$, $\delta=1$. Right column has $\beta=0.05$, $\gamma=0.05$, $\delta=0.05$. Susceptible, infected, and recovered populations are shown on top, middle, and bottom rows, respectively. The legend applies to all figures. All rounding method comparisons were conducted at a time step of 20 PN time steps per unit time per 1 ODE time interval.}
\label{fig:SIRPNfittoODE}
\end{figure}

  In all of the Figure \ref{fig:SIRPNfittoODE}'s subfigures, the PN time steps per unit time ($\tau$) is set to 20. Initially, when viewing Figures \ref{fig:Rounding_1_Sus}, \ref{fig:Rounding_1_Inf}, and \ref{fig:Rounding_1_Rec}, we can see that the non-residual rounding methods are unable to capture the dynamics of the ODE at all. Granted, this is for the most extreme case for the parameter values, but it is important that the PN model is capable of capturing the dynamics for all values [0,1] for $\beta$,$\gamma$, and $\delta$. When looking at more biologically plausible values of $\beta,\gamma,\delta=0.05$ in Figures \ref{fig:Rounding_05_Sus}, \ref{fig:Rounding_05_Inf}, and \ref{fig:Rounding_05_Rec} we can see that the nature of PN firing only when transitions are enabled combined with the extreme value situation as laid out in section \ref{sec:Deal_ex}, the dynamics behave like a 2-cycle. Without the extra logic statements laid out in section \ref{sec:Deal_ex}, the dynamics are even less biologically plausible, though, with no new infected even with high levels of infected and susceptible, simply with $S<\beta S I$.

\subsection{Rescaling of Base Time Step in GPenSIM Model}
\begin{figure}[H]
    \centering
    \textbf{1 GPenSIM Time Step Per ODE Time Unit, $\tau=1$}\par
    \includegraphics[width=1\textwidth]{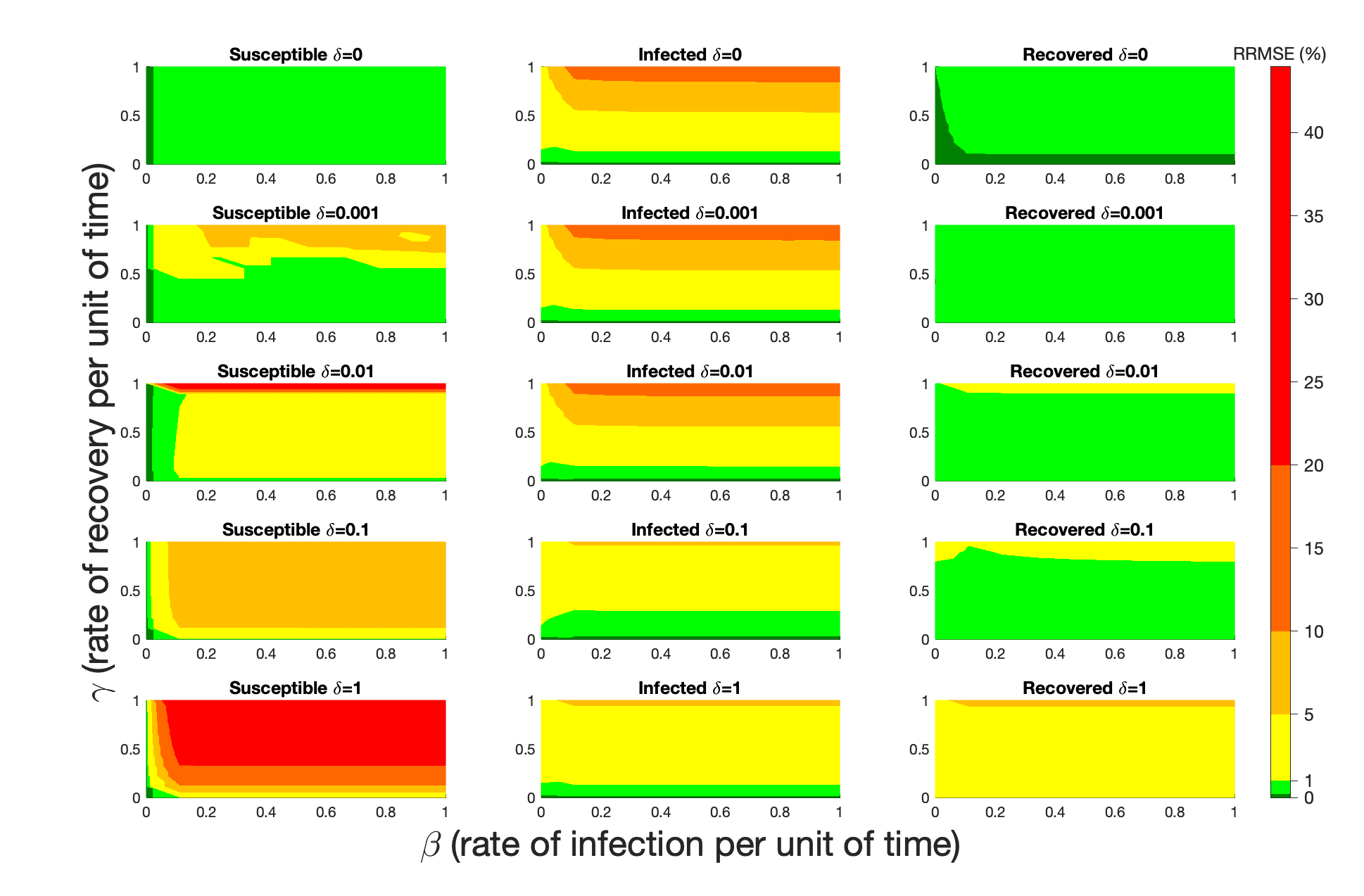}
    \caption{RRMSE percentage across parameter ranges of $[0,1]$ for each respective parameter with $\gamma$ the y-axis of each subfigure, $\beta$ the x-axis of each subfigure, and $\delta$ set at a different fixed value for each subfigure. PN Time Step Per Unit of Time parameter $\tau=1$. Note that red is RRMSE $\leq44\%$ (43.75\% being the max observed RRMSE across all simulations), dark orange is RRMSE $\leq20\%$, light orange is RRMSE $\leq10\%$, yellow is RRMSE $\leq5\%$, light green is RRMSE $\leq1\%$, and dark green is RRMSE $\leq.1\%$}
    \label{fig:Pn_v_ODe_1_T}
\end{figure}

\begin{figure}[H]
    \centering
    \textbf{20 GPenSIM Time Steps Per ODE Time Unit, $\tau=20$}\par
    \includegraphics[width=1\textwidth]{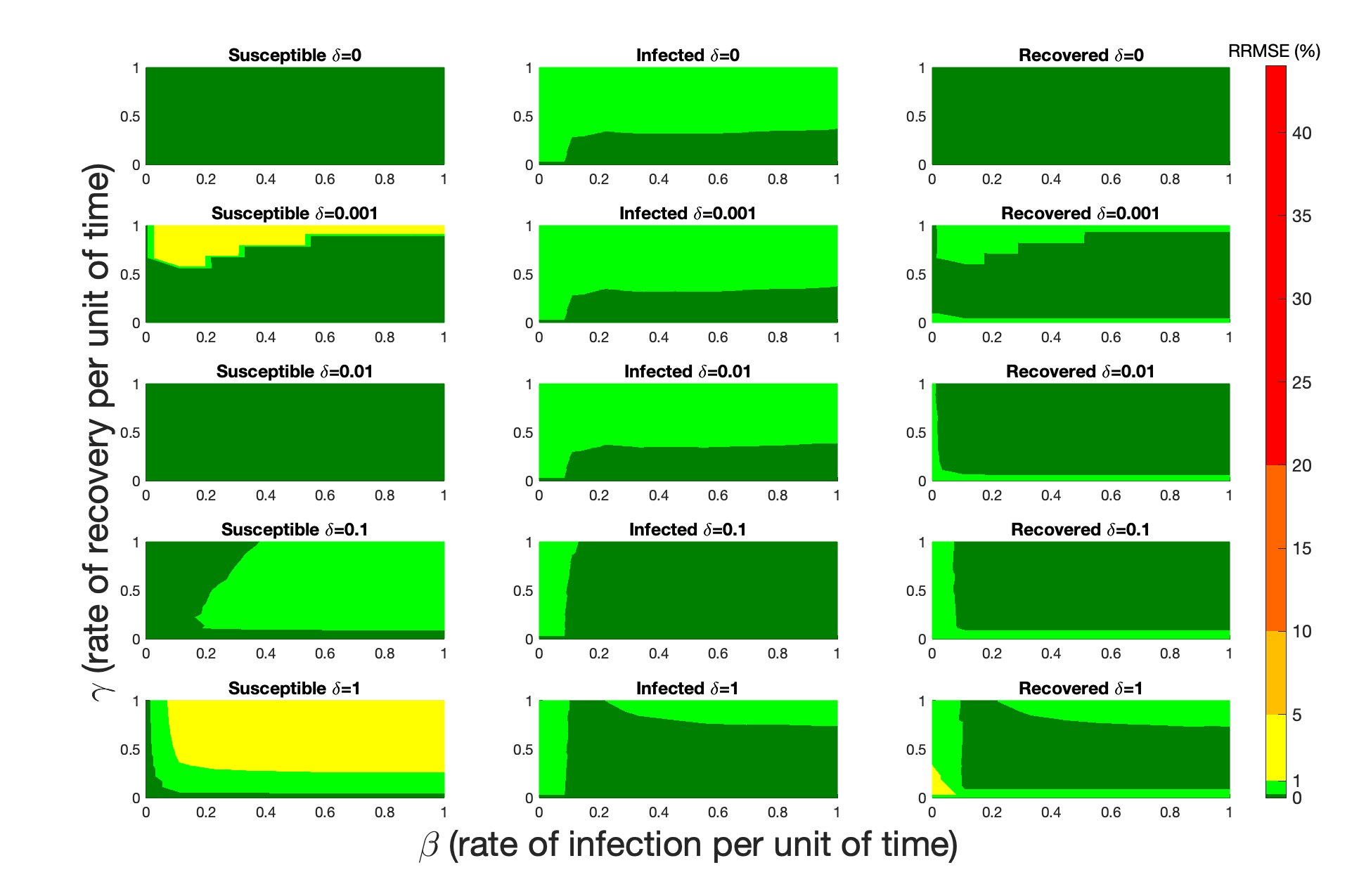}
    \caption{RRMSE percentage across parameter ranges of $[0,1]$ for each respective parameter with $\gamma$ the y-axis of each subfigure, $\beta$ the x-axis of each subfigure, and $\delta$ set at a different fixed value for each subfigure. PN Time Step Per Unit of Time parameter $\tau=20$. Note that yellow is RRMSE $\leq5\%$, light green is RRMSE $\leq1\%$, and dark green is RRMSE $\leq0.1\%$.}
    \label{fig:Pn_v_ODe_20_T}
\end{figure}

\begin{figure}[H]
    \centering
    \textbf{60 GPenSIM Time Steps Per ODE Time Unit, $\tau=60$}\par
    \includegraphics[width=1\textwidth]{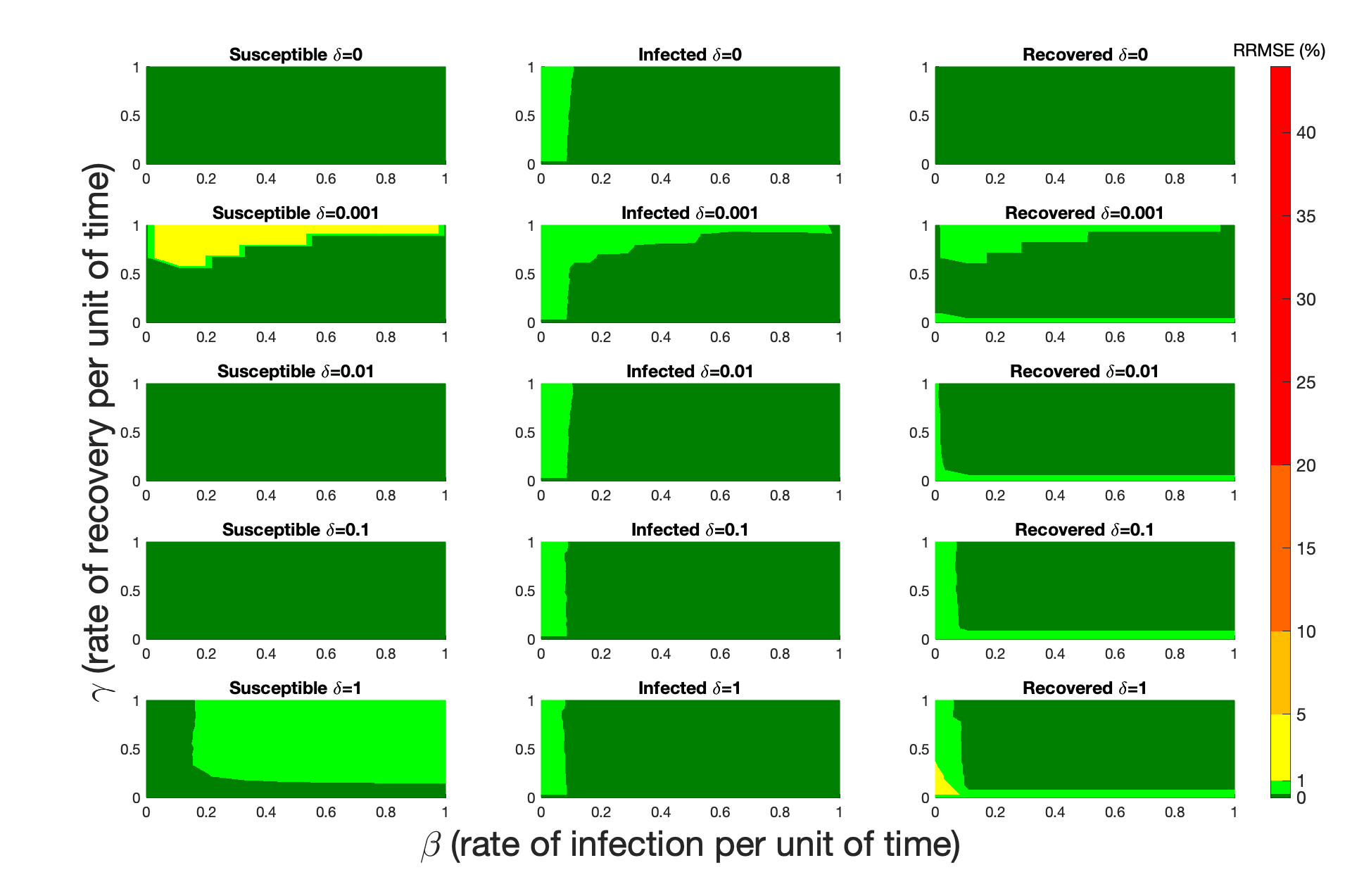}
    \caption{RRMSE percentage across parameter ranges of $[0,1]$ for each respective parameter with $\gamma$ the y-axis of each subfigure, $\beta$ the x-axis of each subfigure, and $\delta$ set at a different fixed value for each subfigure. PN Time Step Per Unit of Time parameter $\tau=60$. Note that yellow is RRMSE $\leq5\%$, light green is RRMSE $\leq1\%$, and dark green is RRMSE $\leq0.1\%$.}
    \label{fig:Pn_v_ODe_60_T}
\end{figure}

\begin{figure}[H]
    \centering
    \includegraphics[width=1\textwidth]{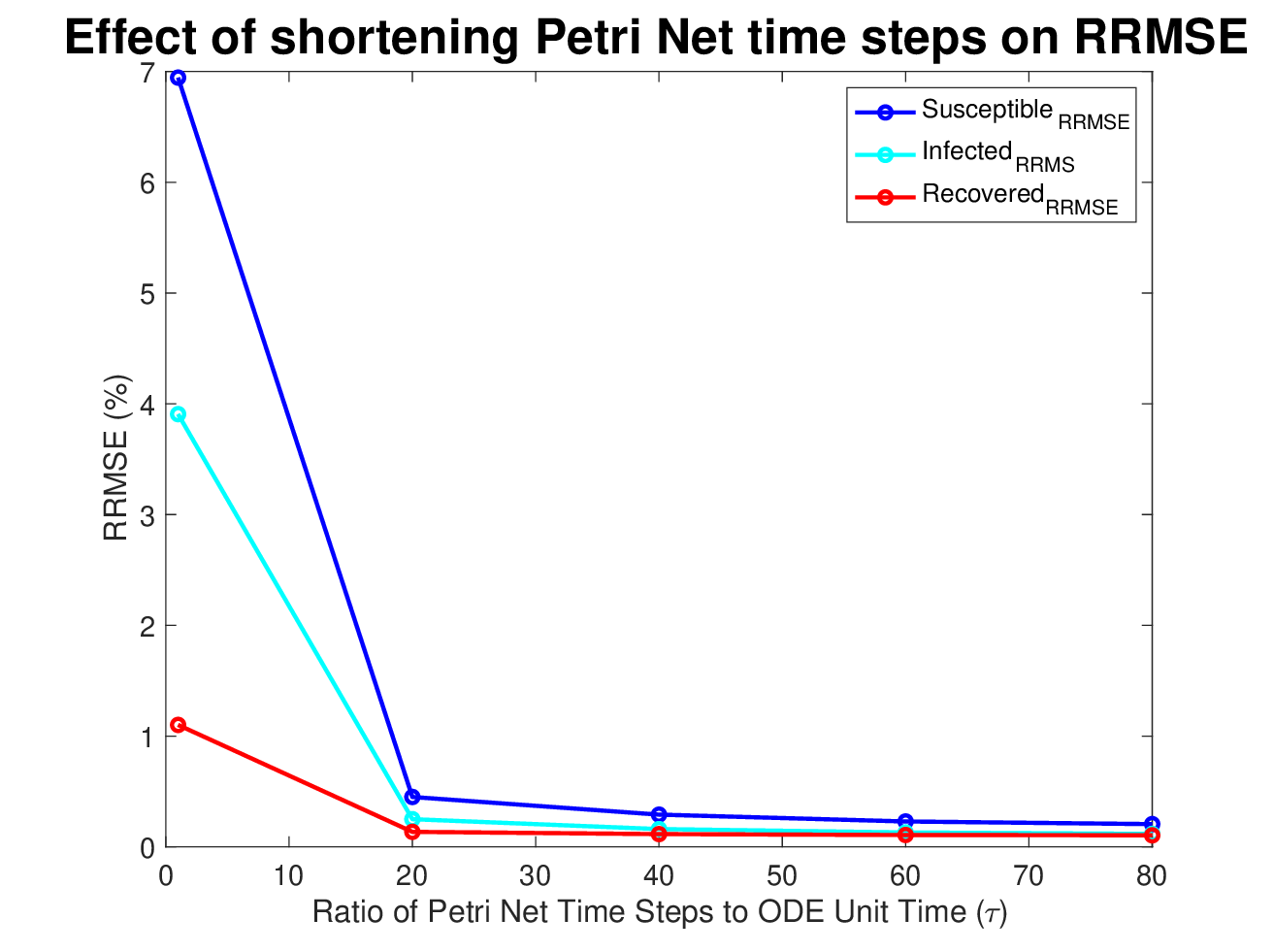}
    \caption{Mean RRMSE in percentage across parameter ranges of $\gamma$ range of [0,1] with 10 linearly spaced points, $\beta$ range of [0,1] with 10 linearly spaced points, and $\delta$ range of[0,1] with 5 logarithmically spaced points.}
    \label{fig:SIR_PN_ODE_fit}
\end{figure}

\begin{figure}[H]
    \centering
    \includegraphics[width=1\textwidth]{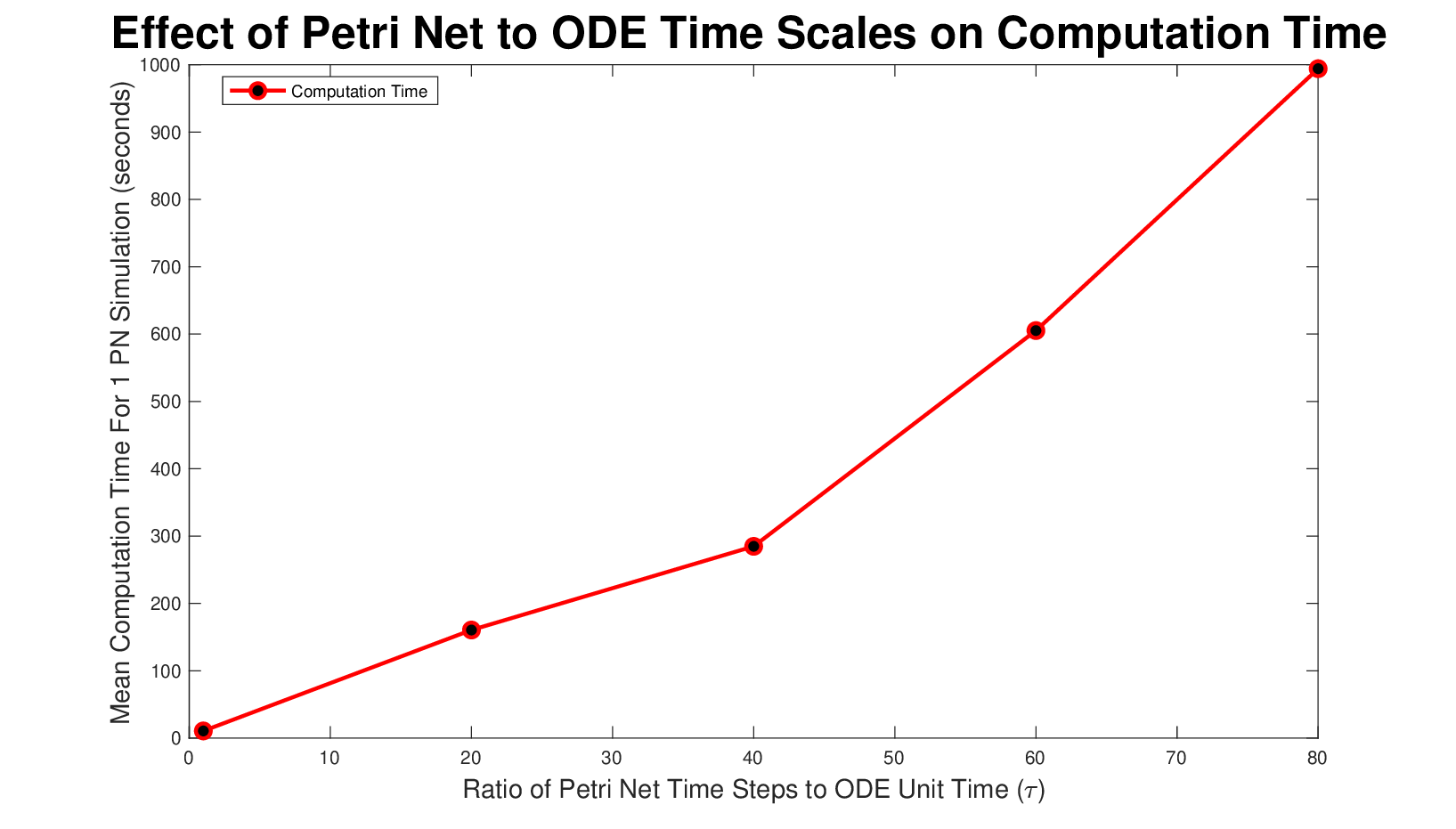}
    \caption{Mean computation time for one dynamic arc weight PN model run in GPenSIM for various times steps as seen in Figures \ref{fig:Pn_v_ODe_1_T},\ref{fig:Pn_v_ODe_20_T},\ref{fig:Pn_v_ODe_40_T}, \ref{fig:Pn_v_ODe_60_T}, and \ref{fig:Pn_v_ODe_80_T}.}
    \label{fig:Comp_Time}
\end{figure}

\begin{figure}[H]
    \centering
    \textbf{GPenSIM Variable Arc Weight PN Population Scalar=2, $\tau=60$}\par
    \includegraphics[width=1\textwidth]{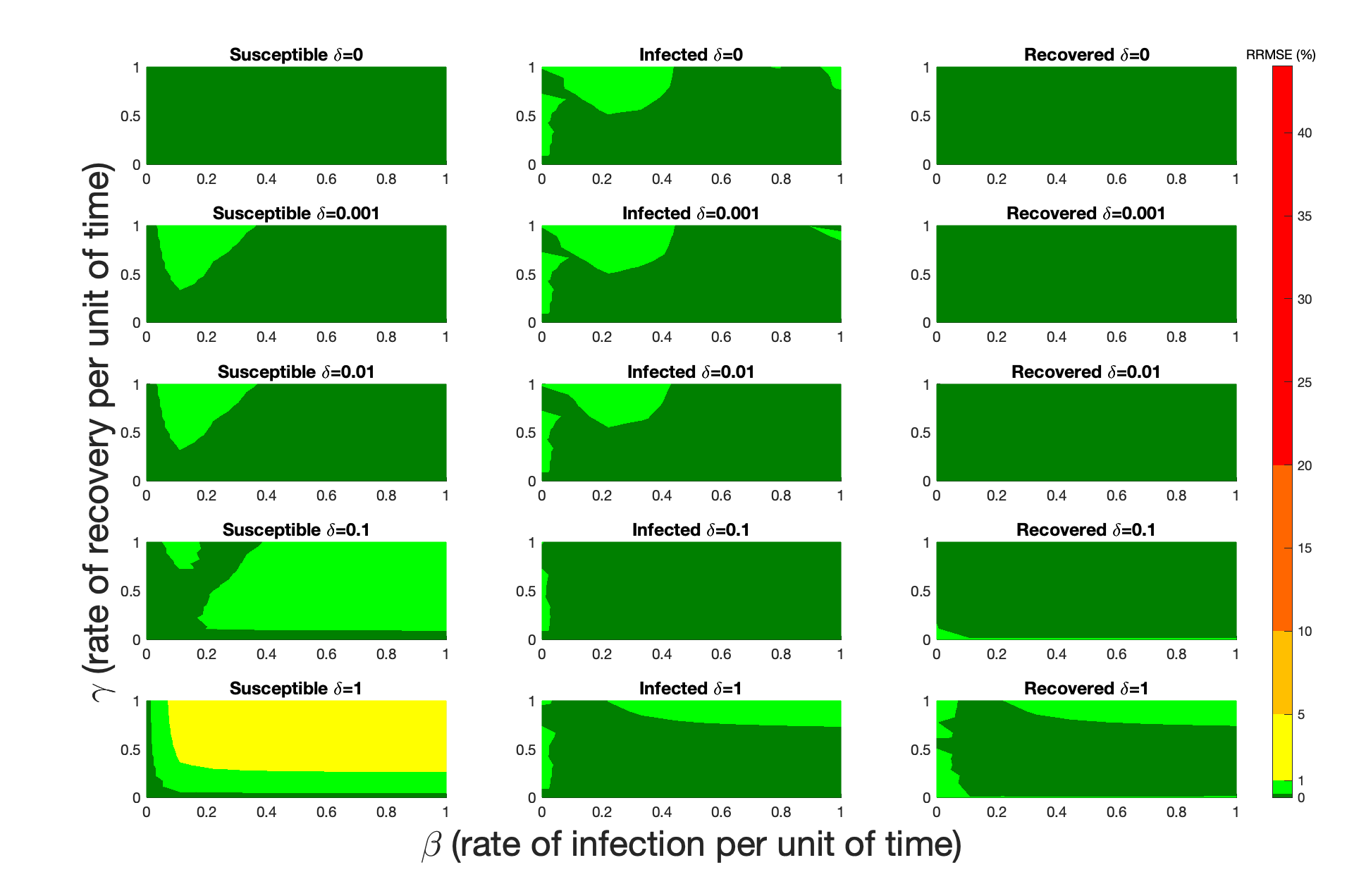}
    \caption{RRMSE percentage across parameter ranges of $[0,1]$ for each respective parameter with $\gamma$ the y-axis of each subfigure, $\beta$ the x-axis of each subfigure, and $\delta$ set at a different fixed value for each subfigure. PN Time Step Per Unit of Time parameter $\tau=60$. Note that yellow is RRMSE$\leq5\%$, light green is RRMSE $\leq1\%$, and dark green is RRMSE $\leq.1\%$}
    \label{fig:PopScalar_2_tau_60}
\end{figure}

\begin{figure}[H]
    \centering
    \textbf{GPenSIM Variable Arc Weight PN Population Scalar=5, $\tau=60$}\par
    \includegraphics[width=1\textwidth]{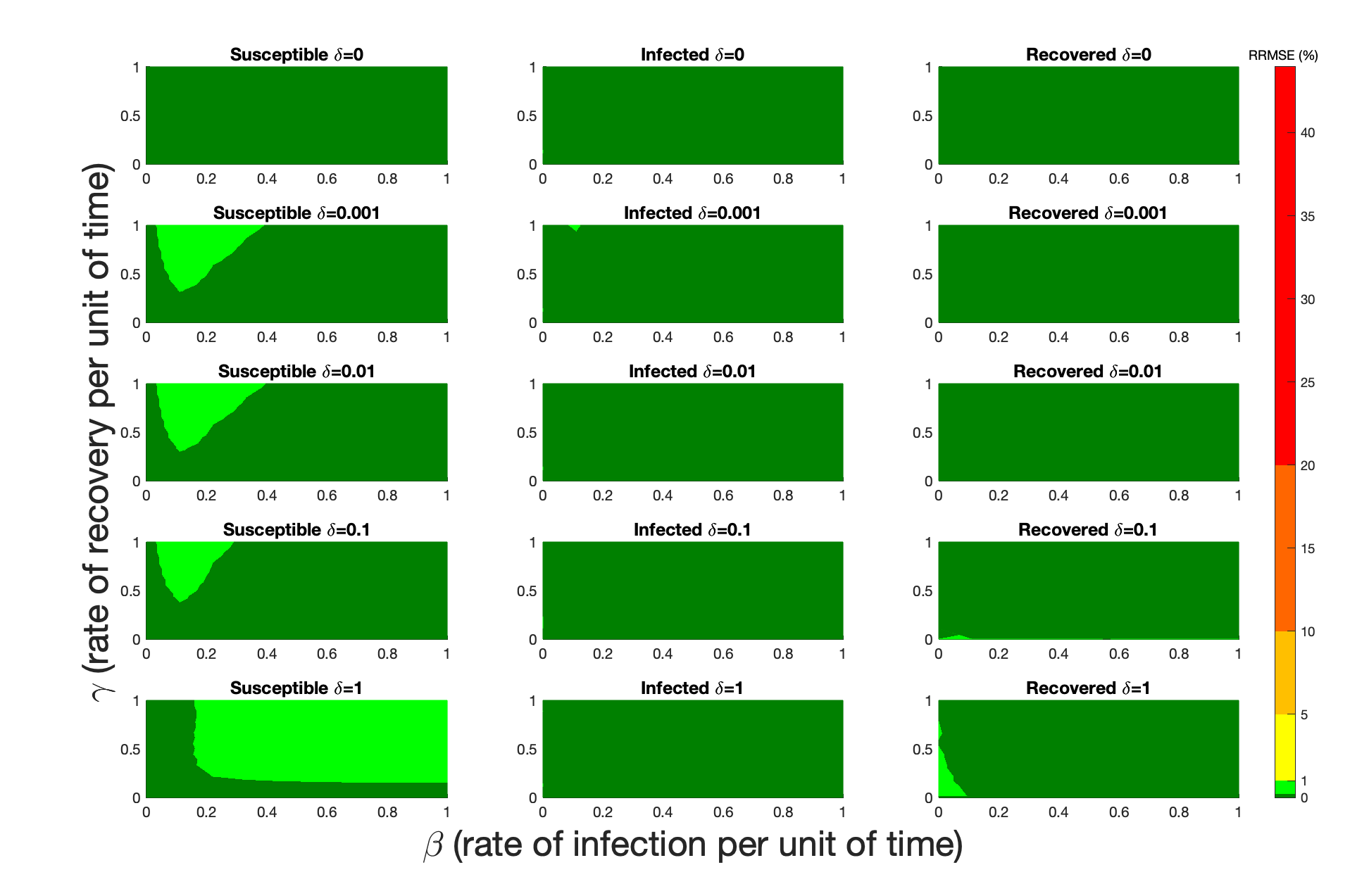}
    \caption{RRMSE percentage across parameter ranges of $[0,1]$ for each respective parameter with $\gamma$ the y-axis of each subfigure, $\beta$ the x-axis of each subfigure, and $\delta$ set at a different fixed value for each subfigure. PN Time Step Per Unit of Time parameter $\tau=60$. Note that light green is RRMSE $\leq1\%$, and dark green is RRMSE $\leq.1\%$}
    \label{fig:PopScalar_5_tau_60}
\end{figure}

\begin{figure}[H]
    \centering
    \textbf{Spike Tau Leaping Stochastic PN, Population Scalar=1}\par
    \includegraphics[width=1\textwidth]{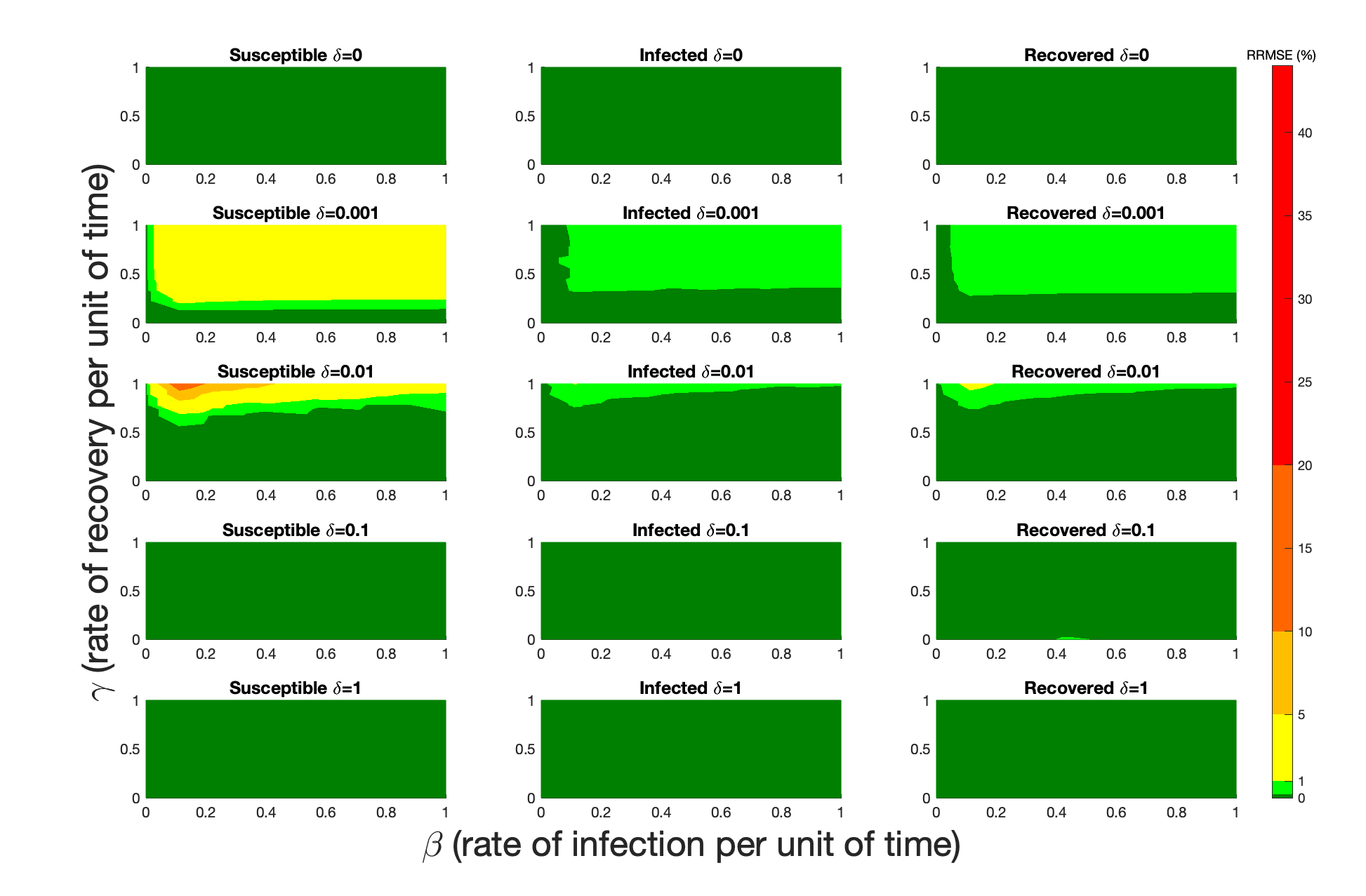}
    \caption{RRMSE percentage across parameter ranges of $[0,1]$ for each respective parameter with $\gamma$ the y-axis of each subfigure, $\beta$ the x-axis of each subfigure, and $\delta$ set at a different fixed value for each subfigure. Note that red is RRMSE $\leq44\%$ (43.75\% being the max observed RRMSE across all simulations), dark orange is RRMSE $\leq20\%$, light orange is RRMSE $\leq10\%$, yellow is RRMSE $\leq5\%$, light green is RRMSE $\leq1\%$, and dark green is RRMSE $\leq.1\%$}
    \label{fig:tauLeapingpop1}
\end{figure}

\begin{figure}[H]
    \centering
    \textbf{Spike Tau Leaping Stochastic PN, Population Scalar=10}\par
    \includegraphics[width=1\textwidth]{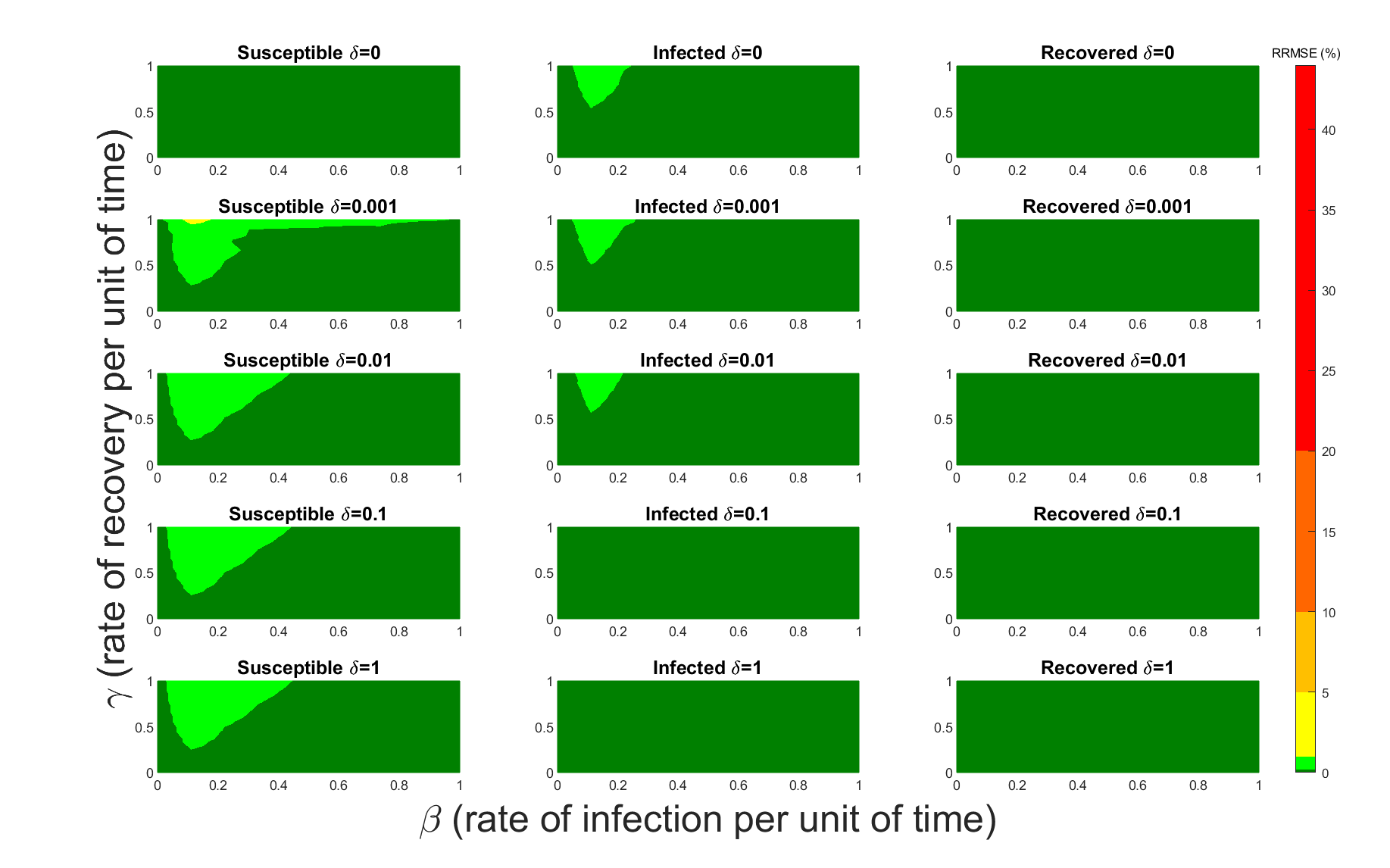}
    \caption{RRMSE percentage across parameter ranges of $[0,1]$ for each respective parameter with $\gamma$ the y-axis of each subfigure, $\beta$ the x-axis of each subfigure, and $\delta$ set at a different fixed value for each subfigure. Note that yellow is RRMSE $\leq5\%$, light green is RRMSE $\leq1\%$, and dark green is RRMSE $\leq.1\%$}
    \label{fig:tauLeapingpop10}
\end{figure}

\begin{figure}[H]
    \centering
    \textbf{Spike Tau Leaping Direct Stochastic PN, Population Scalar=100}\par
    \includegraphics[width=1\textwidth]{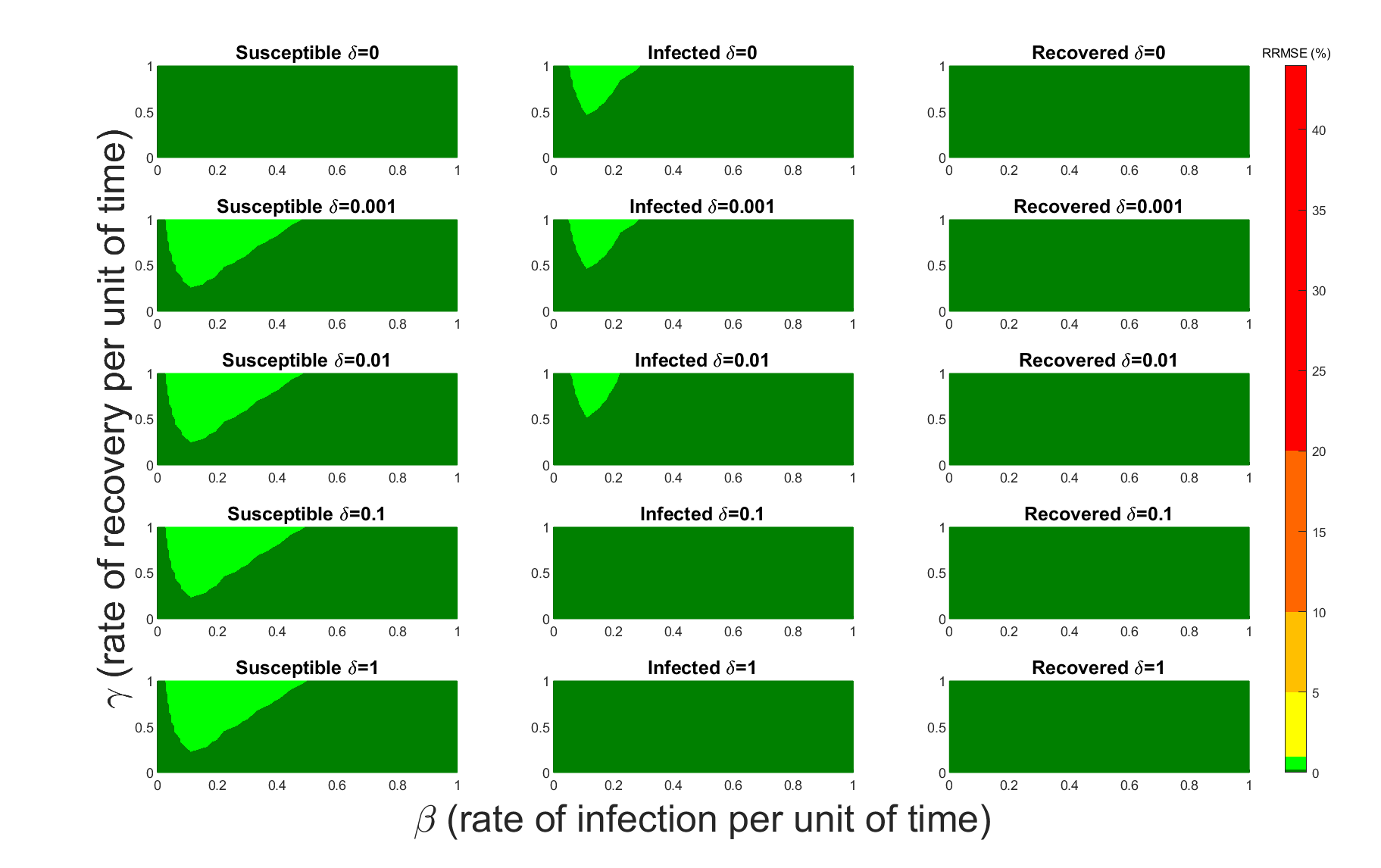}
    \caption{RRMSE percentage across parameter ranges of $[0,1]$ for each respective parameter with $\gamma$ the y-axis of each subfigure, $\beta$ the x-axis of each subfigure, and $\delta$ set at a different fixed value for each subfigure. Note that light green is RRMSE $\leq1\%$, and dark green is RRMSE $\leq.1\%$}
    \label{fig:tauLeapingpop100}
\end{figure}

\begin{figure}[H]
    \centering
    \textbf{Forty PN Time Steps Per Unit of Time, $\tau=40$}\par
    \includegraphics[width=1\textwidth]{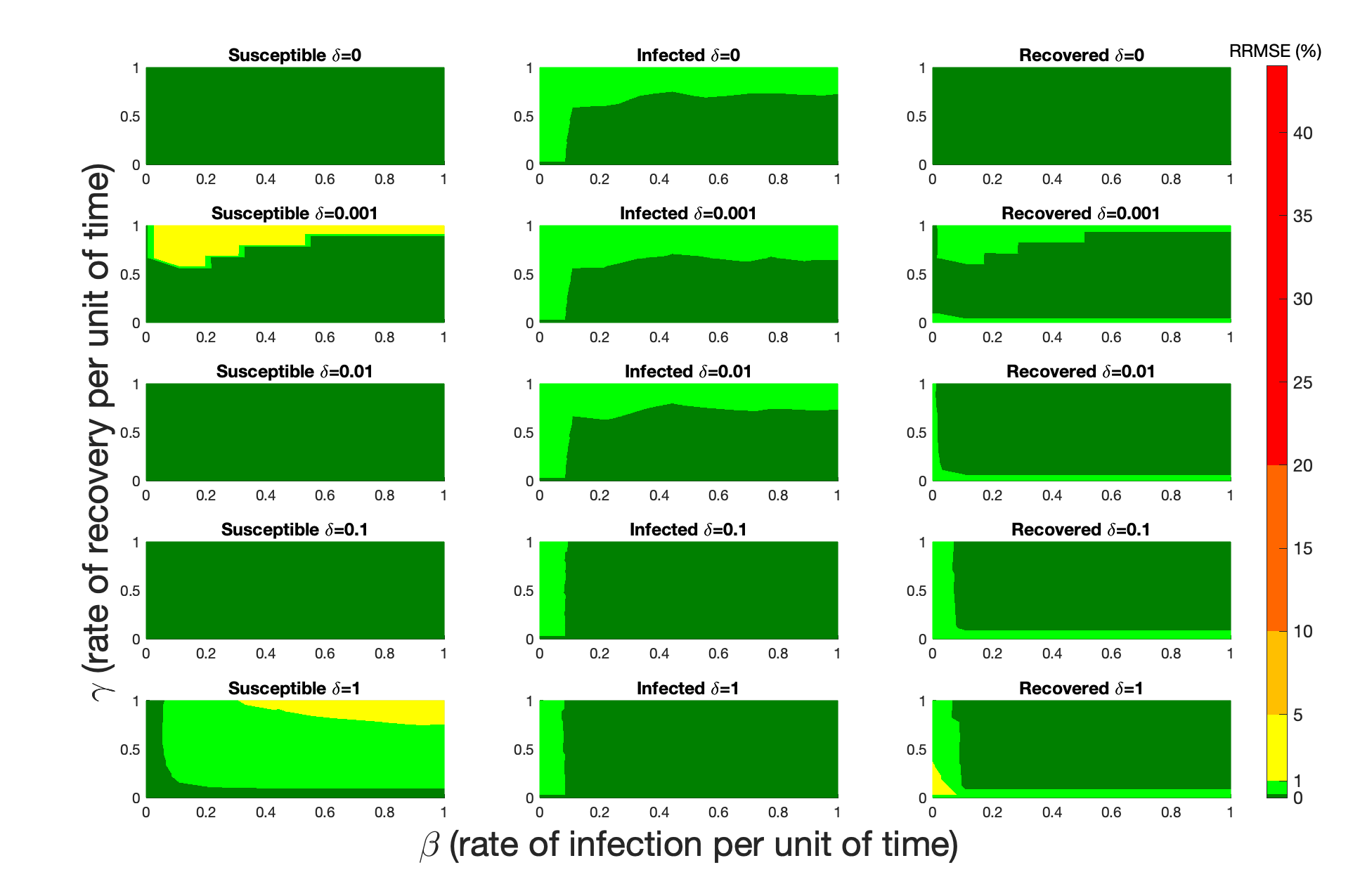}
    \caption{RRMSE percentage across parameter ranges of $[0,1]$ for each respective parameter with $\gamma$ the y-axis of each subfigure, $\beta$ the x-axis of each subfigure, and $\delta$ set at a different fixed value for each subfigure. PN Time Step Per Unit of Time parameter $\tau=40$. Note that yellow is RRMSE $\leq5\%$, light green is RRMSE $\leq1\%$, and dark green is RRMSE $\leq0.1\%$.}
    \label{fig:Pn_v_ODe_40_T}
\end{figure}

\begin{figure}[H]
    \centering
    \textbf{Eighty PN Time Steps Per Unit of Time, $\tau=80$}\par
    \includegraphics[width=1\textwidth]{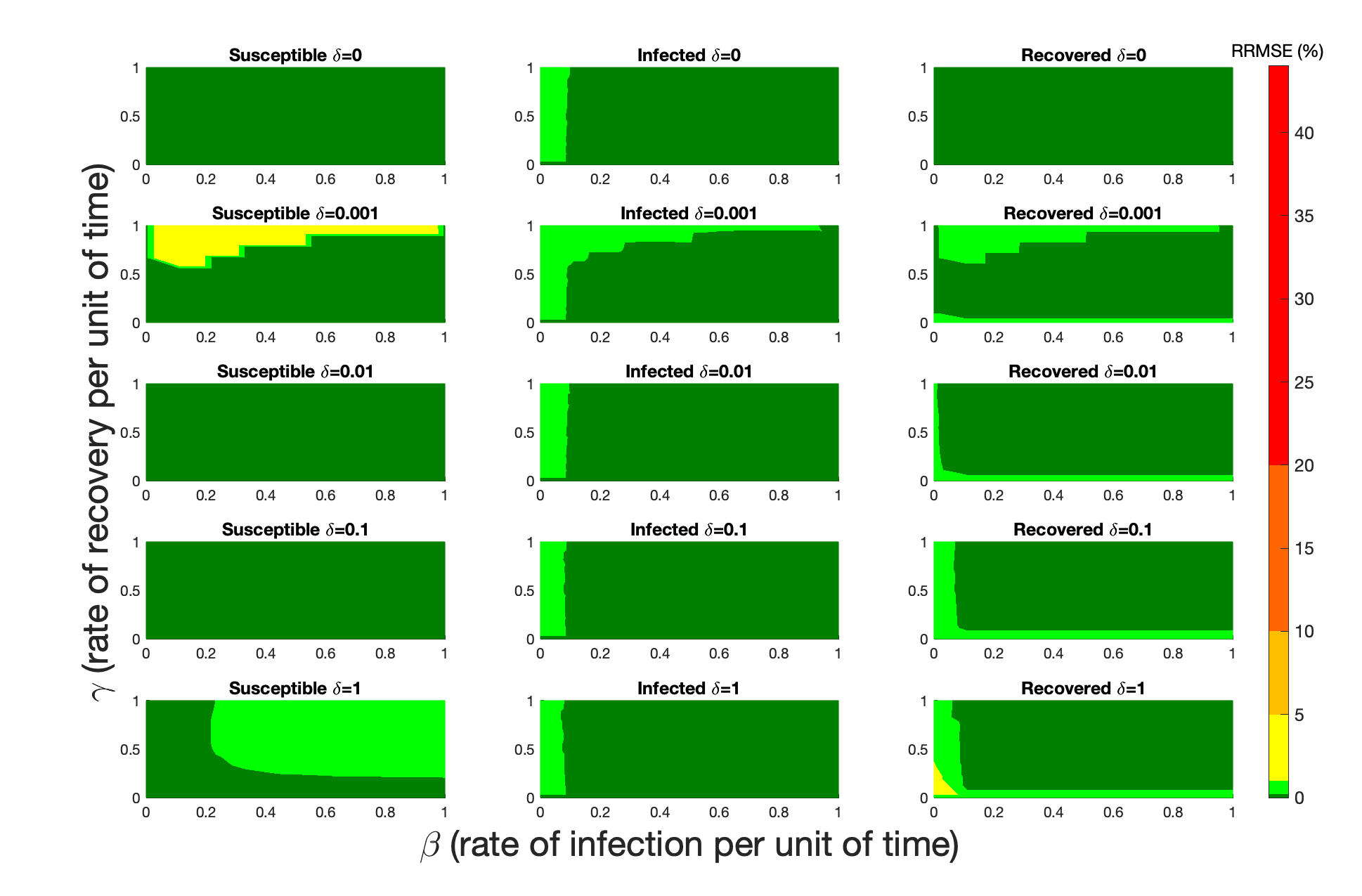}
    \caption{RRMSE percentage across parameter ranges of $[0,1]$ for each respective parameter with $\gamma$ the y-axis of each subfigure, $\beta$ the x-axis of each subfigure, and $\delta$ set at a different fixed value for each subfigure. PN Time Step Per Unit of Time parameter $\tau=80$. Note that yellow is RRMSE $\leq5\%$, light green is RRMSE $\leq1\%$, and dark green is RRMSE $\leq0.1\%$.}
    \label{fig:Pn_v_ODe_80_T}
\end{figure}

The PN model, as seen in Figure \ref{fig:SIRS_PN}, with different PN time steps per unit time ($\tau$) are compared to the ODE model Equations \ref{eq:S_SIRS}-\ref{eq:R_SIRS} with the same parameters values using RRMSE. The parameter grid chosen was a linearly spaced grid of size ten between [0,1] for parameters $\beta$ and $\gamma$. These are the x-axis and y-axis, respectively, for all of the sub-figures in Figures \ref{fig:Pn_v_ODe_1_T}-\ref{fig:Pn_v_ODe_80_T}. Then for $\delta$ there is a logarithmic spaced grid of size five between [0,1] going from the top to bottom row of Figures \ref{fig:Pn_v_ODe_1_T}-\ref{fig:Pn_v_ODe_80_T}.

When the $\tau=1$, in Figure \ref{fig:Pn_v_ODe_1_T} the RRMSE performance is relatively poor, especially for higher values of $\beta$, $\gamma$, and $\delta$. This low $\tau$ value displays the problem of comparing a discrete versus a continuous time system with large swings in population happening instantly at the time step, not allowing the dynamics of the PN to come close to matching that of an ODE continuous system. These sharp dynamics combined with the additional firing mechanisms of PN means the RRMSE values produced are relatively large. 

Figure \ref{fig:Pn_v_ODe_20_T} shows the vast improvement in RRMSE with utilizing higher PN time steps per unit time. With this one change, the maximum RRMSE becomes approximately $4.3$, and the visual improvement at extreme values is considerable.

As $\tau$ increases from one to eighty, there is an overall reduction of RRMSE with each subsequent increase of the PN time steps per unit time, which can be more clearly seen in Figure \ref{fig:SIR_PN_ODE_fit}. This can be seen for each run in Figures \ref{fig:Pn_v_ODe_1_T},\ref{fig:Pn_v_ODe_20_T},\ref{fig:Pn_v_ODe_60_T} and Figures \ref{fig:Pn_v_ODe_40_T},\ref{fig:Pn_v_ODe_80_T} in the appendix.


\subsection{Population Scaling for GPenSIM and Spike models} 
\label{sec:pop_scal}

The stochastic Petri Net system used in Spike and the dynamic arc weight Petri Net system used in GPenSIM both use discrete time and rounding of arcs weights to whole numbers. As such, it was theorized that scaling the initial populations and descaling after the simulations have occurred would reduce overall RRMSE \cite{beccuti_analysis_2014}. 

\begin{figure}[H]
    \centering
    \textbf{GPenSIM Variable Arc Weight PN Population Scalar=4, $\tau=60$}\par
    \includegraphics[width=1\textwidth]{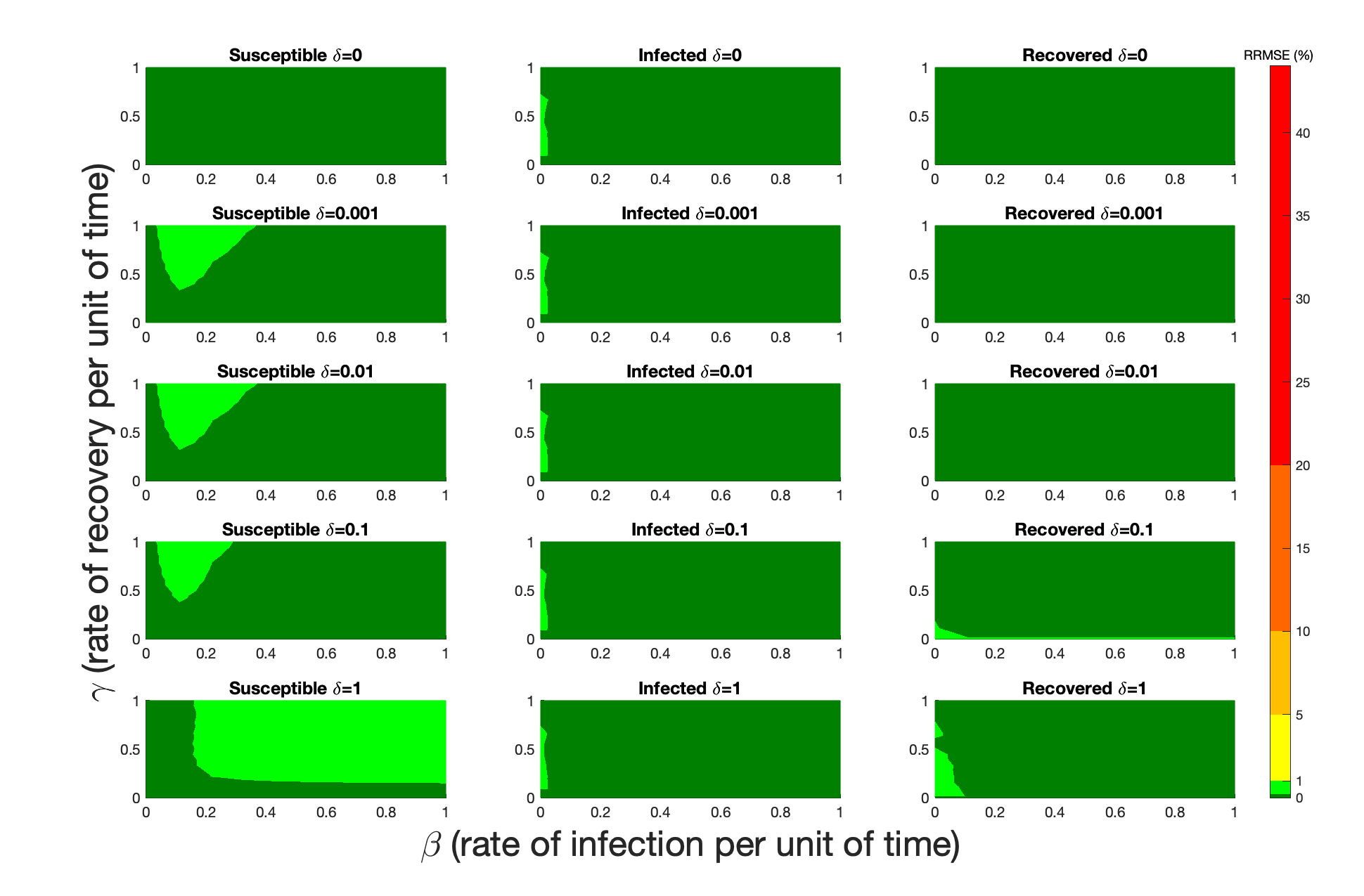}
    \caption{RRMSE percentage across parameter ranges of $[0,1]$ for each respective parameter with $\gamma$ the y-axis of each subfigure, $\beta$ the x-axis of each subfigure, and $\delta$ set at a different fixed value for each subfigure. PN Time Step Per Unit of Time parameter $\tau=60$. Note that light green is RRMSE $\leq1\%$, and dark green is RRMSE $\leq.1\%$}
    \label{fig:PopScalar_4_tau_60}
\end{figure}

For GPenSIM we started with a $\tau=60$ since this setting lead to small RRMSEs overall and lower computation time. Figure \ref{fig:PopScalar_4_tau_60} shows the results of rescaling the population size by a factor of 4.


In Figure \ref{fig:PopScalar_4_tau_60}, all of the parameter value combinations produce RRMSE less than 1\%, with the majority of the values being less than 0.1\%. While it depends on the application, we consider this to be an acceptable level for the Petri Net model.

\begin{figure}[H]
    \centering
    \includegraphics[width=1\textwidth]{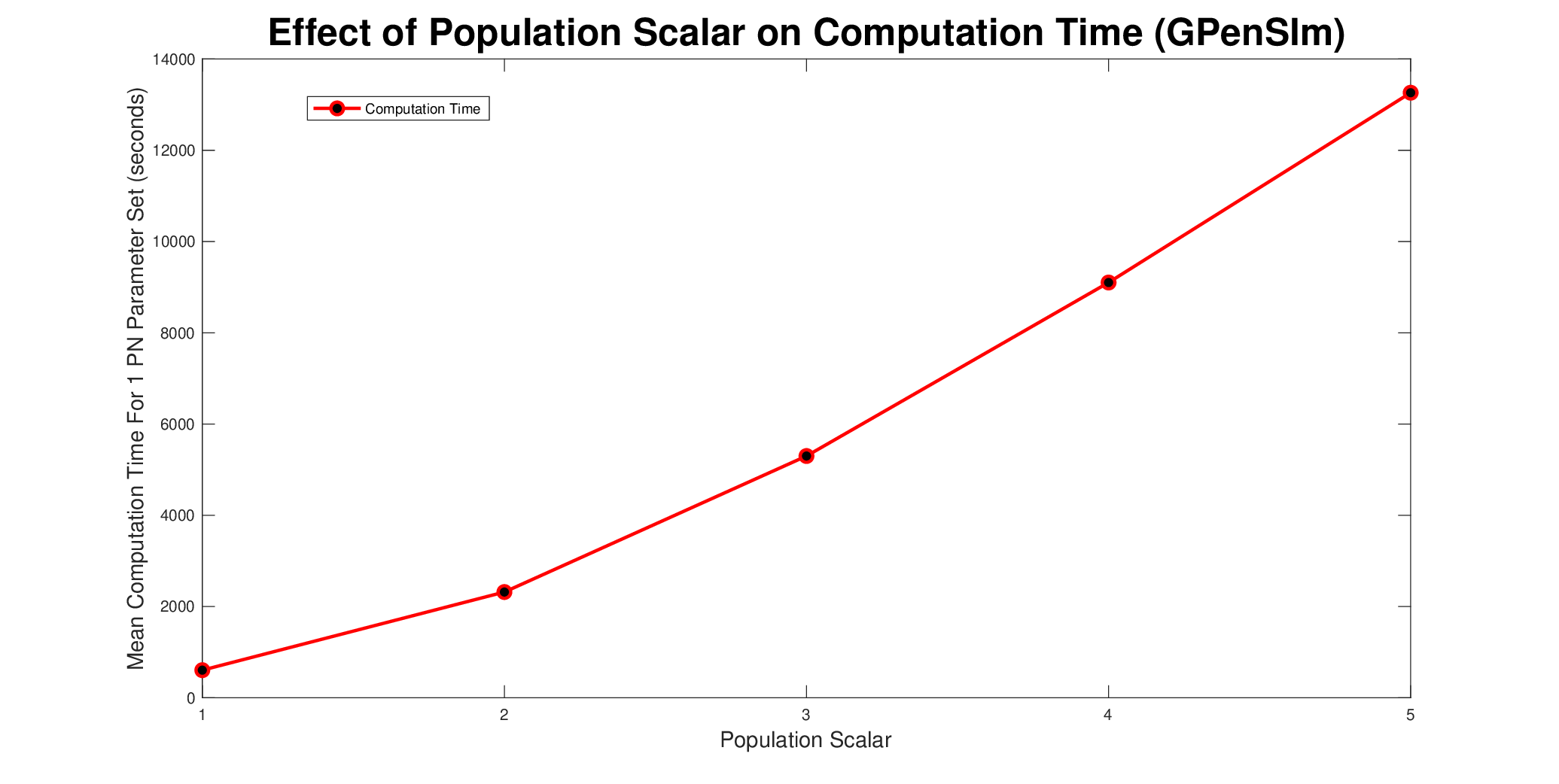}
    \caption{Mean computation time for one dynamic arc weight PN model run in GPenSIM for increasing population scalars as seen in Figures \ref{fig:Pn_v_ODe_60_T},\ref{fig:PopScalar_2_tau_60},\ref{fig:PopScalar_4_tau_60}, and \ref{fig:PopScalar_5_tau_60}.}
    \label{fig:Comp_Time_popscalar_GPenSIM}
\end{figure}

For Spike, we can not directly alter the $\tau$ value, but we can directly alter the initial population size. Given Spike's relatively low computation time for the SPN, we were able to raise the population scalar to much higher values, starting at one and going up by order of magnitude each time. We stopped when we reached a level of less than 1\% RRMSE across all parameter values for the direct and tau-leaping method for SPNs. Though Spike demonstrates significantly better computational efficiency relative to GPenSIM, we limited all SPN simulations in Spike to 500 trials. This ensures manageable computation times while maintaining the accuracy and statistical robustness needed for meaningful results.

\begin{figure}[H]
    \centering
    \textbf{Spike Direct Stochastic PN, Population Scalar=1}\par
    \includegraphics[width=1\textwidth]{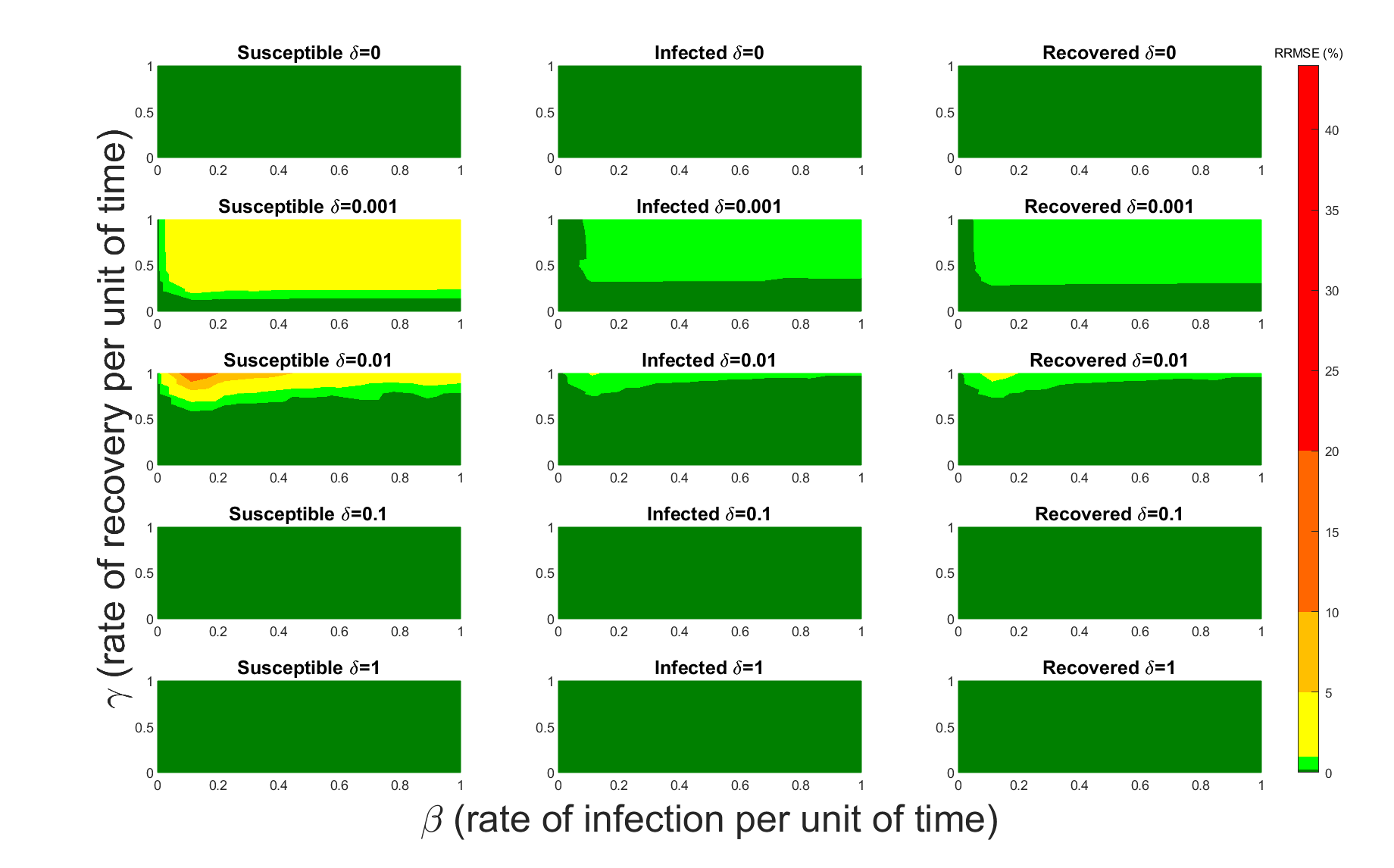}
    \caption{RRMSE percentage across parameter ranges of $[0,1]$ for each respective parameter with $\gamma$ the y-axis of each subfigure, $\beta$ the x-axis of each subfigure, and $\delta$ set at a different fixed value for each subfigure. Note that red is RRMSE $\leq44\%$ (43.75\% being the max observed RRMSE across all simulations), dark orange is RRMSE $\leq20\%$, light orange is RRMSE $\leq10\%$, yellow is RRMSE $\leq5\%$, light green is RRMSE $\leq1\%$, and dark green is RRMSE $\leq.1\%$}
    \label{fig:directpop1}
\end{figure}

\begin{figure}[H]
    \centering
    \textbf{Spike Direct Stochastic PN, Population Scalar=10}\par
    \includegraphics[width=1\textwidth]{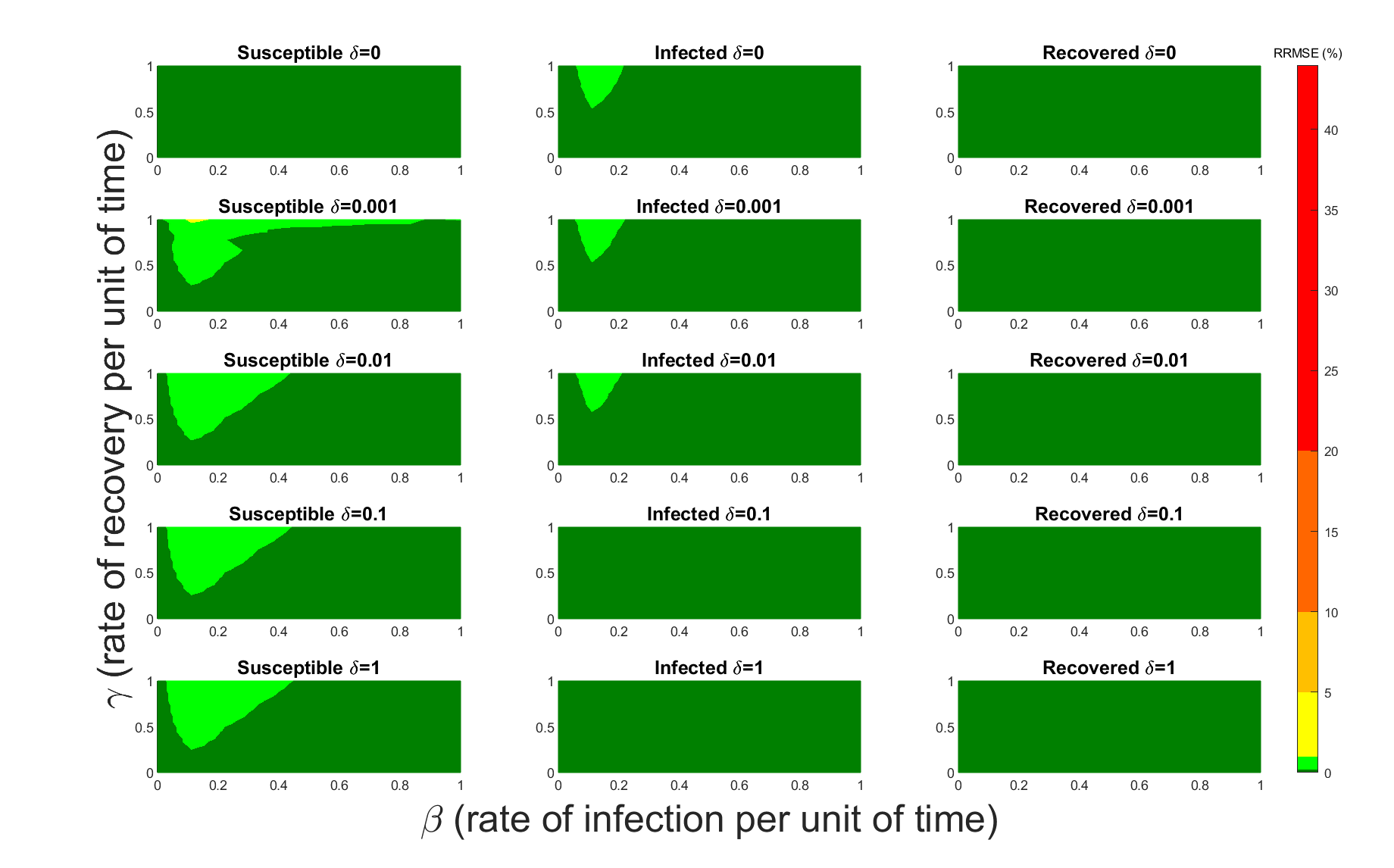}
    \caption{RRMSE percentage across parameter ranges of $[0,1]$ for each respective parameter with $\gamma$ the y-axis of each subfigure, $\beta$ the x-axis of each subfigure, and $\delta$ set at a different fixed value for each subfigure. Note that yellow is RRMSE $\leq5\%$, light green is RRMSE $\leq1\%$, and dark green is RRMSE $\leq.1\%$}
    \label{fig:directpop10}
\end{figure}

\begin{figure}[H]
    \centering
    \textbf{Spike Direct Stochastic PN, Population Scalar=100}\par
    \includegraphics[width=1\textwidth]{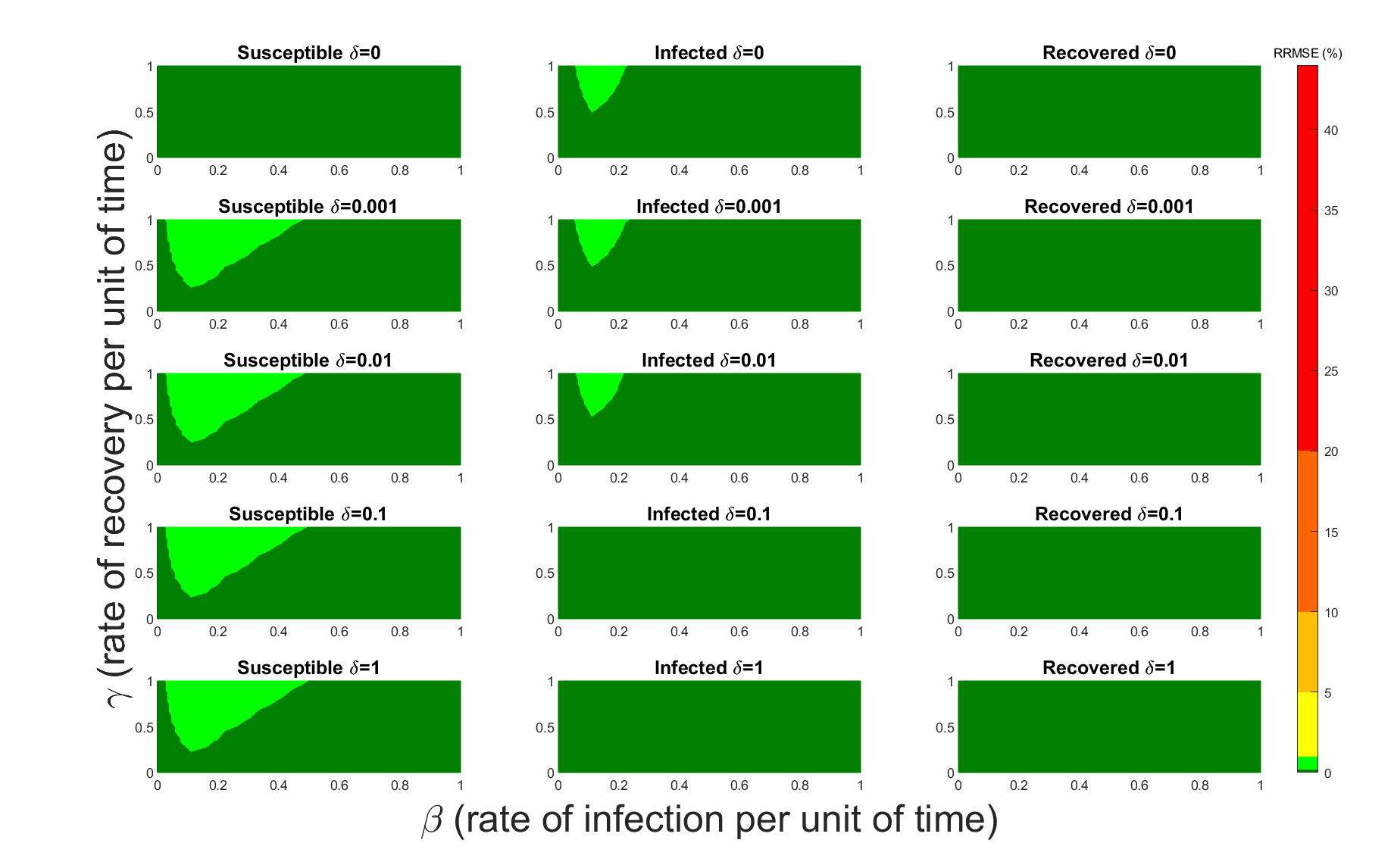}
    \caption{RRMSE percentage across parameter ranges of $[0,1]$ for each respective parameter with $\gamma$ the y-axis of each subfigure, $\beta$ the x-axis of each subfigure, and $\delta$ set at a different fixed value for each subfigure. Note that light green is RRMSE $\leq1\%$, and dark green is RRMSE $\leq.1\%$}
    \label{fig:directpop100}
\end{figure}

\begin{figure}[H]
    \centering
    \includegraphics[width=1\textwidth]{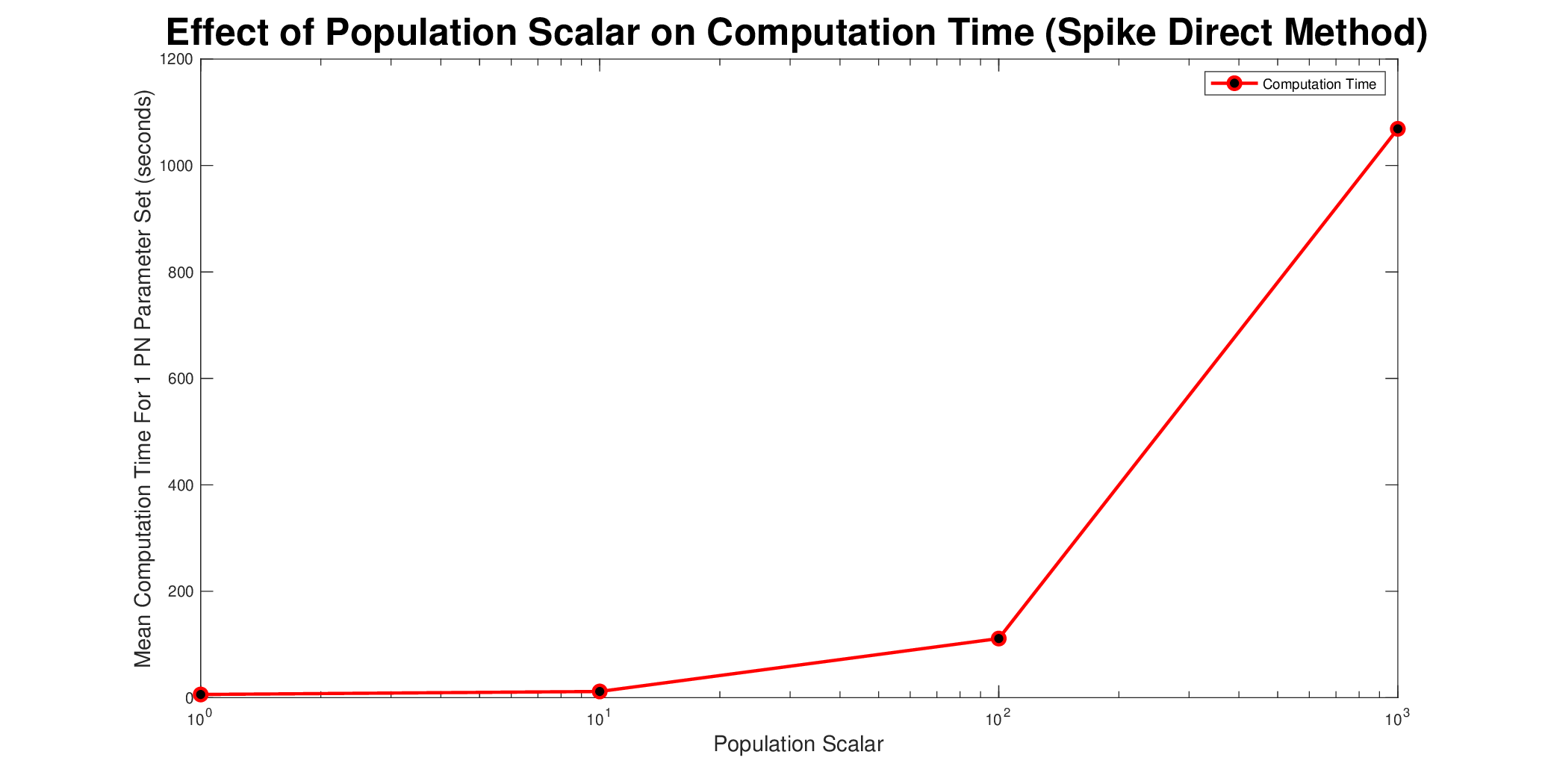}
    \caption{Mean computation time for one dynamic arc weight PN model run in Spike using the Direct method for increasing population scalars as seen in Figures \ref{fig:directpop1}-\ref{fig:directpop100}.}
    \label{fig:Comp_Time_popscalar_Spk_Direct}
\end{figure}

In Figures \ref{fig:directpop1}-\ref{fig:directpop100} we see the RRMSE drop as the population scalar increases, to a level where for all parameter values are less than 1\% RRMSE with a population scalar of 100 for the direct method.

\begin{figure}[H]
    \centering
    \includegraphics[width=1\textwidth]{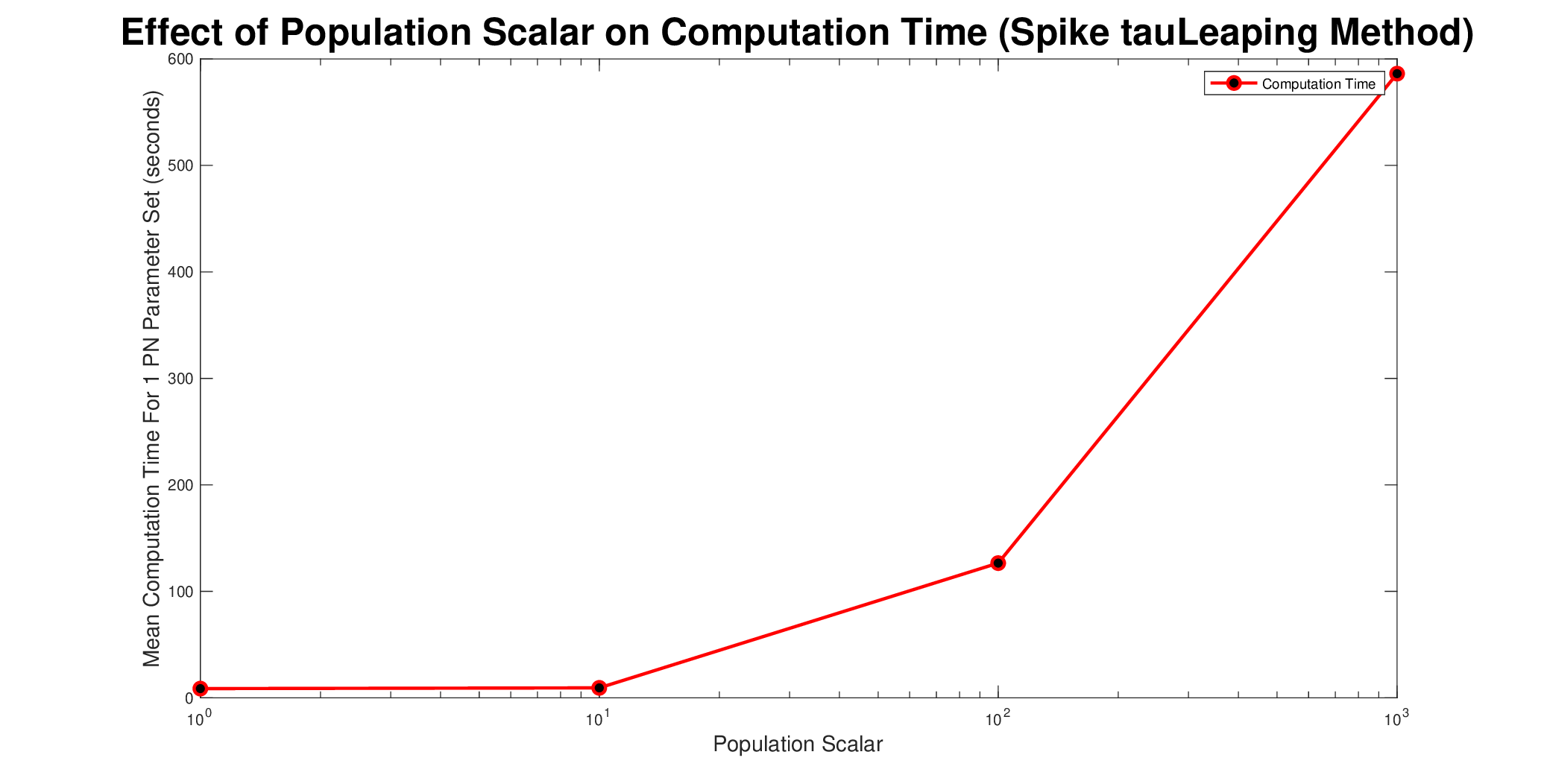}
    \caption{Mean computation time for one dynamic arc weight PN model run in Spike using the tauLeaping method for increasing population scalars as seen in Figures \ref{fig:tauLeapingpop1}-\ref{fig:tauLeapingpop100}.}
    \label{fig:Comp_Time_popscalar_Spk_tauL}
\end{figure}

In Figures \ref{fig:directpop1}-\ref{fig:directpop100} we see the RRMSE drop as the population scalar increases, to a level where for all parameter values are less than 1\% RRMSE with a population scalar of 100 for the direct method. Beyond a population scalar of 100, the regions of less than 0.1\% RRMSE remain nearly identical when the population scalar is raised by an order of magnitude. This likely indicates that either an asymptotic convergence of the means is occurring or some other factor is limiting the the RRMSE from decreasing further.

We note that the computing demand for single runs of these simulations are low enough to run on a laptop and do not require a supercomputer cluster. However, when running large sets of parameters, with large population scalars and large $\tau$ values parallelization, a supercomputer, or both, will allow for the simulations to be run in a reasonable time. For Figures \ref{fig:Comp_Time},\ref{fig:Comp_Time_popscalar_GPenSIM},\ref{fig:Comp_Time_popscalar_Spk_Direct}, and \ref{fig:Comp_Time_popscalar_Spk_tauL} the values were found while being run in MATLAB 2024b, using GPenSIM v.10 software and Spike v1.6 respectively, on a computer with a 2.3 GHz 8-Core processor, with 64 GB or memory. These models were also run on the Arizona State University SOL supercomputer, where even when allocated four cores and 32 GB of memory, the models only had a maximum memory used value of 5.6 GB.

\section{Discussion and Conclusions}\label{sec:conlusion}

Petri Nets are increasingly used to model the dynamics of infectious disease spread, building on a deep literature of ODE models. However, few studies have examined the numerical equivalency of Petri Net models to reference ODE systems, especially beyond the most basic SIR model. In this paper, we introduced a novel PN implementation of SIR-type models using a discrete-time, variable arc weight framework in the software package GPenSIM. We explored the numerical equivalency of this variable arc weight model to an SIRS ODE model, and we found that several choices in implementation are important for matching the ODE dynamics, including the rounding method used for non-integer arc weights and rescaling of the time step for large arc weight values. We also compared the behavior of a continuous-time Markov Chain PN using the software package Spike. For both types of PN, we found that rescaling the population size led to major improvements in the relative root mean square error comparing the PN trajectories to the reference ODE behavior. This indicates that PN models do not necessarily generate identical dynamics to their ODE counterparts, and that prior numerical studies are an important tool for ensuring similarity.

Variable arc weight Petri Nets using GPenSIM software represent a useful technical breakthrough in Petri Net modeling by allowing dynamic adjustments to the arc weights based on real-time conditions or external state changes rather than static values. GPenSIM's ability to incorporate dynamic arc weights yields an additional PN modeling option beyond the CTMC framework. However, we found that the code we implemented for the variable arc weight models showed worse computational efficiency than CTMC models at similar levels of performance matching the ODE. GPenSIM was designed for modeling pure discrete-event systems, whereas this project involves converting a continuous system into a discrete one before applying Petri Nets. 
GPenSIM instead is designed to encourage users to break models into smaller, modular Petri Modules and distribute their execution across multiple CPUs for efficiency. Since a general goal for PN researchers is to construct large-scale, complex models, this suggests that improved efficiency will be an important direction for developing large-scale connected models in the GPenSIM environment.

\section*{Funding}
This research is supported by NIH grant 5R01GM131405-02 encompassing Trevor Reckell and Bright Manu under primary investigators Petar Jevti\'{c} and Beckett Sterner.

\section*{Acknowledgments}

\section*{Conflict of Interest}
The authors declare no conflict of interest.

\clearpage
\bibliography{02_ref}
\bibliographystyle{ieeetr}

\section{Supplementary Index} 

\subsection{Additional $tau$ runs GPenSIM}

\subsection{Rounding Method runs GPenSIM}
\begin{figure}[H]
    \centering
    \textbf{Standard Plus Residual Rounding, $\tau=20$}\par
    \includegraphics[width=1\textwidth]{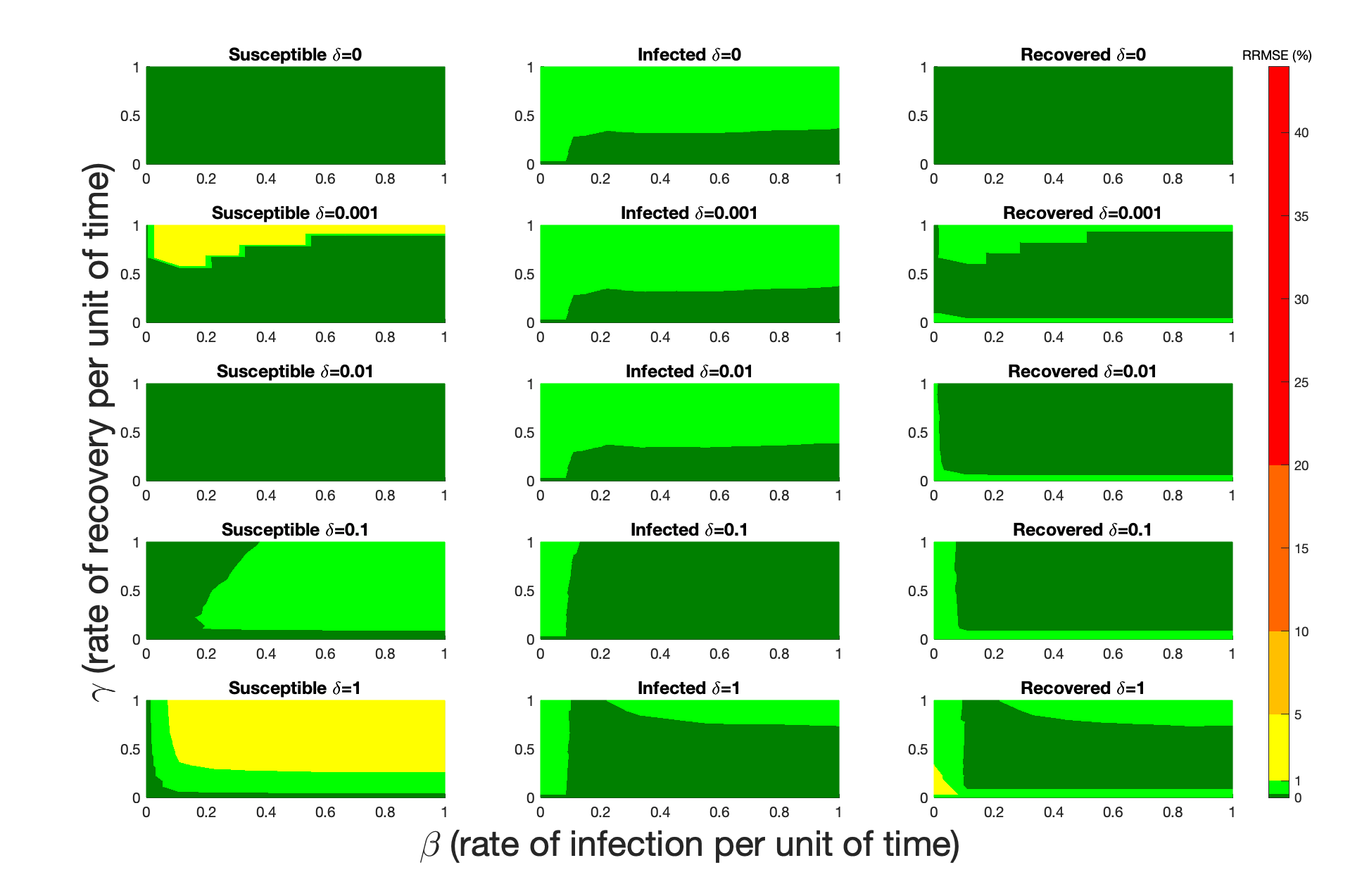}
    \caption{RRMSE percentage across parameter ranges of $[0,1]$ for each respective parameter with $\gamma$ the y-axis of each subfigure, $\beta$ the x-axis of each subfigure, and $\delta$ set at a different fixed value for each subfigure. PN Time Step Per Unit of Time parameter $\tau=1$. Note that red is RRMSE $\leq44\%$ (43.75\% being the max observed RRMSE across all simulations), dark orange is RRMSE $\leq20\%$, light orange is RRMSE $\leq10\%$, yellow is RRMSE $\leq5\%$, light green is RRMSE $\leq1\%$, and dark green is RRMSE $\leq.1\%$}
    \label{fig:Rounding_S+R}
\end{figure}

\begin{figure}[H]
    \centering
    \textbf{Floor Function Rounding, $\tau=20$}\par
    \includegraphics[width=1\textwidth]{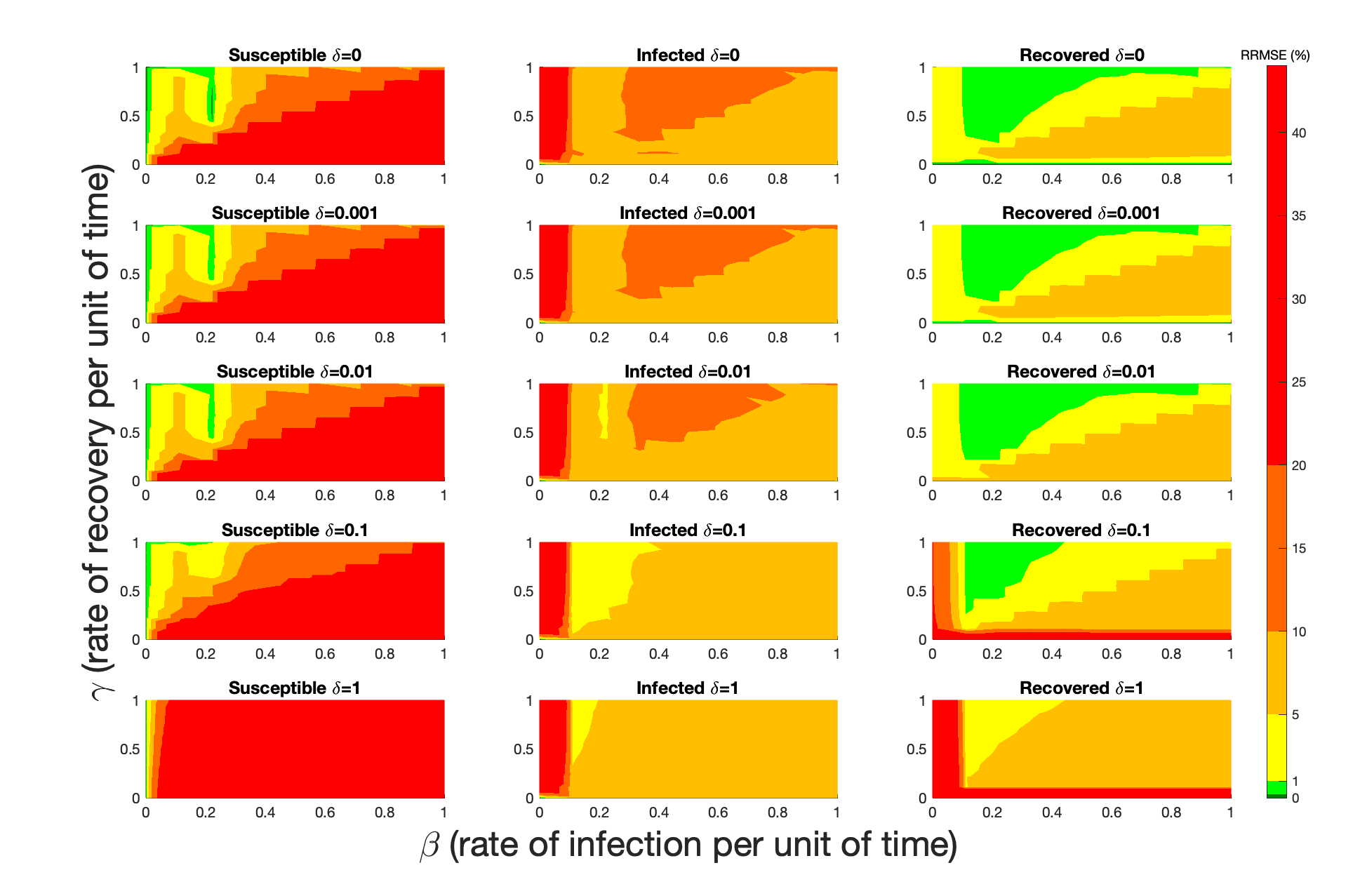}
    \caption{RRMSE percentage across parameter ranges of $[0,1]$ for each respective parameter with $\gamma$ the y-axis of each subfigure, $\beta$ the x-axis of each subfigure, and $\delta$ set at a different fixed value for each subfigure. PN Time Step Per Unit of Time parameter $\tau=1$. Note that red is RRMSE $\leq44\%$ (43.75\% being the max observed RRMSE across all simulations), dark orange is RRMSE $\leq20\%$, light orange is RRMSE $\leq10\%$, yellow is RRMSE $\leq5\%$, light green is RRMSE $\leq1\%$, and dark green is RRMSE $\leq.1\%$}
    \label{fig:Rounding_Floor}
\end{figure}

\begin{figure}[H]
    \centering
    \textbf{Ceiling Function plus Residual, $\tau=20$}\par
    \includegraphics[width=1\textwidth]{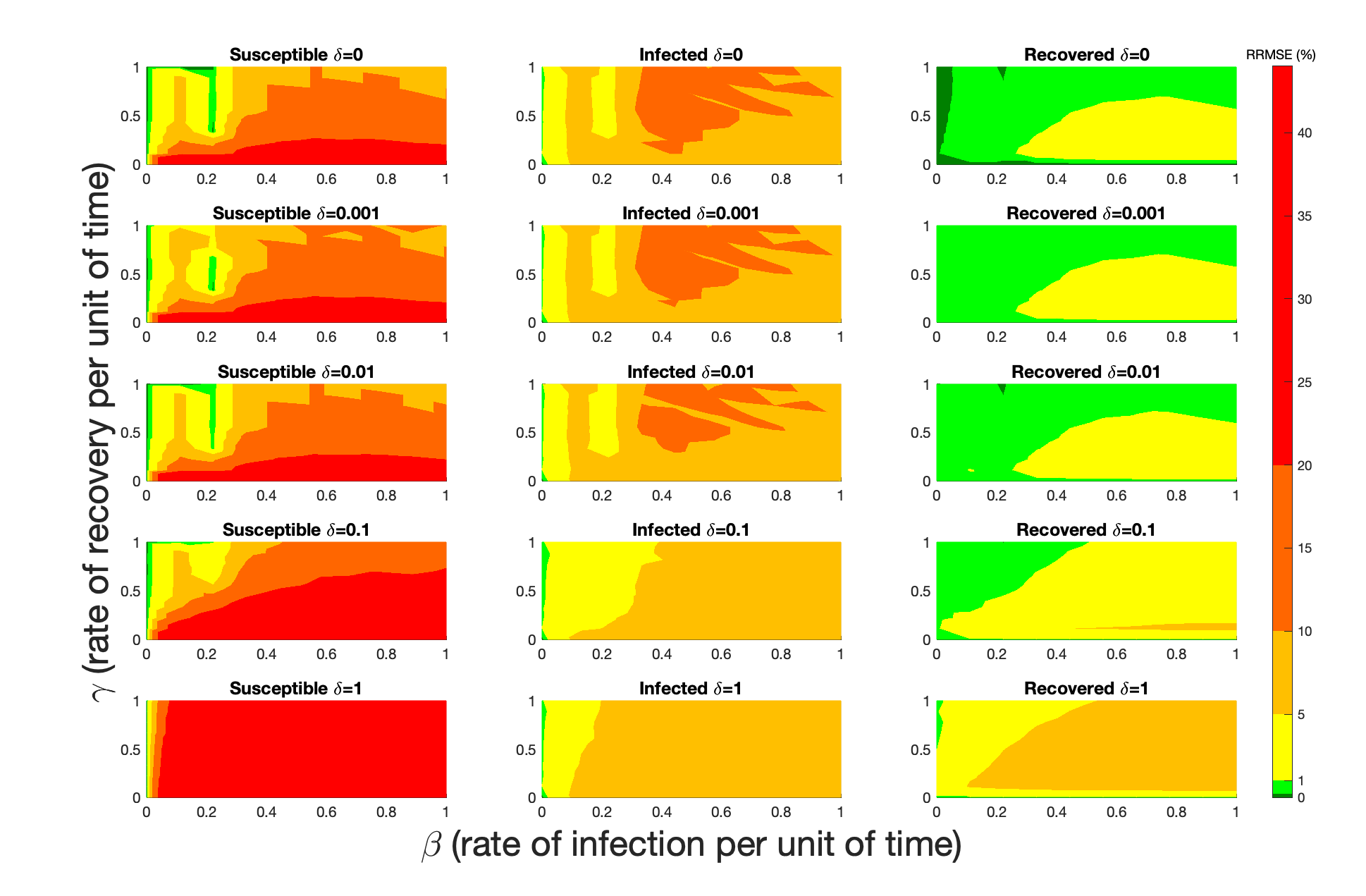}
    \caption{RRMSE percentage across parameter ranges of $[0,1]$ for each respective parameter with $\gamma$ the y-axis of each subfigure, $\beta$ the x-axis of each subfigure, and $\delta$ set at a different fixed value for each subfigure. PN Time Step Per Unit of Time parameter $\tau=1$. Note that red is RRMSE $\leq44\%$ (43.75\% being the max observed RRMSE across all simulations), dark orange is RRMSE $\leq20\%$, light orange is RRMSE $\leq10\%$, yellow is RRMSE $\leq5\%$, light green is RRMSE $\leq1\%$, and dark green is RRMSE $\leq.1\%$}
    \label{fig:Rounding_C+R}
\end{figure}

\begin{figure}[H]
    \centering
    \textbf{Ceiling Function Rounding, $\tau=20$}\par
    \includegraphics[width=1\textwidth]{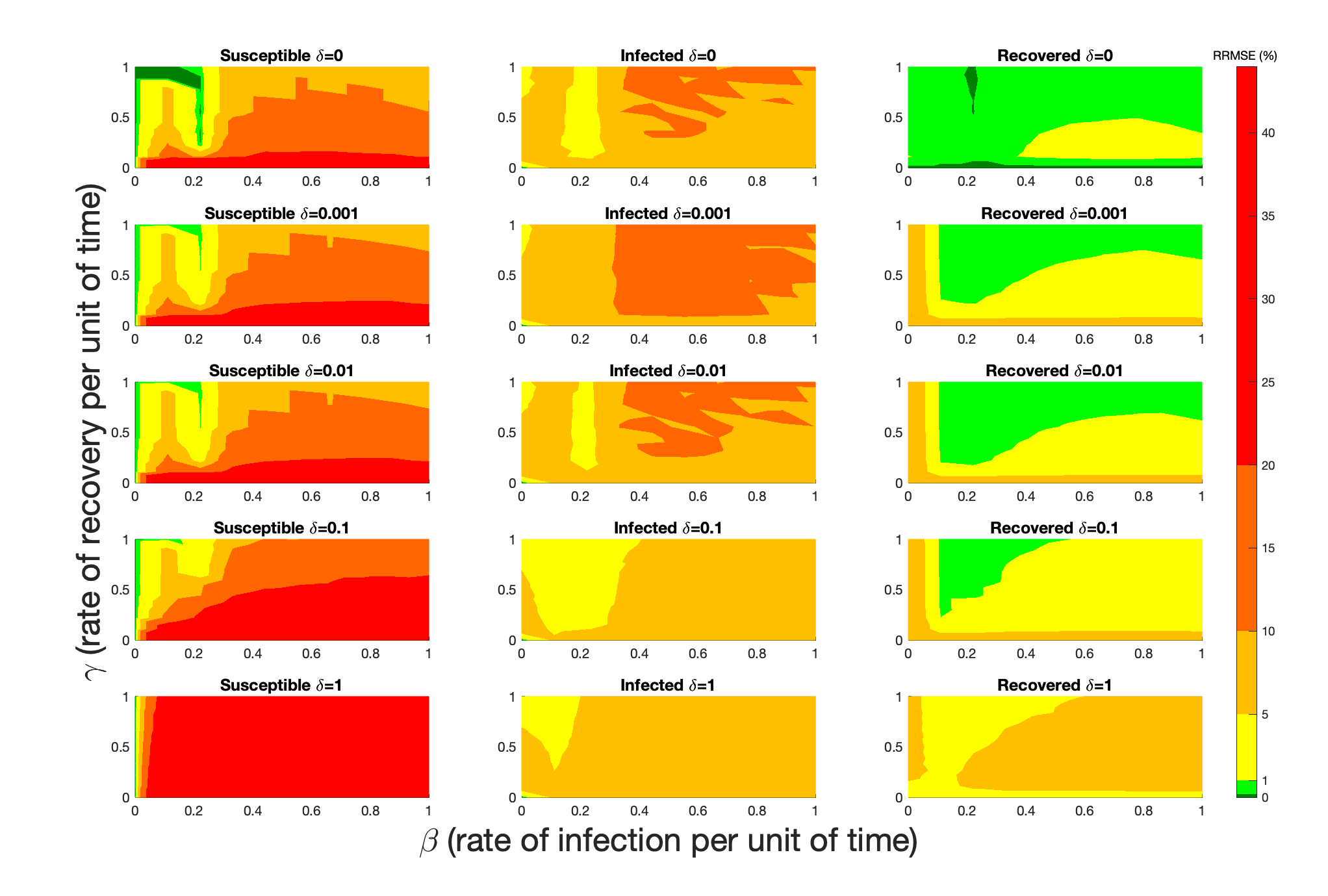}
    \caption{RRMSE percentage across parameter ranges of $[0,1]$ for each respective parameter with $\gamma$ the y-axis of each subfigure, $\beta$ the x-axis of each subfigure, and $\delta$ set at a different fixed value for each subfigure. PN Time Step Per Unit of Time parameter $\tau=1$. Note that red is RRMSE $\leq44\%$ (43.75\% being the max observed RRMSE across all simulations), dark orange is RRMSE $\leq20\%$, light orange is RRMSE $\leq10\%$, yellow is RRMSE $\leq5\%$, light green is RRMSE $\leq1\%$, and dark green is RRMSE $\leq.1\%$}
    \label{fig:Rounding_Ceiling}
\end{figure}

\begin{figure}[H]
    \centering
    \textbf{Standard Rounding, $\tau=20$}\par
    \includegraphics[width=1\textwidth]{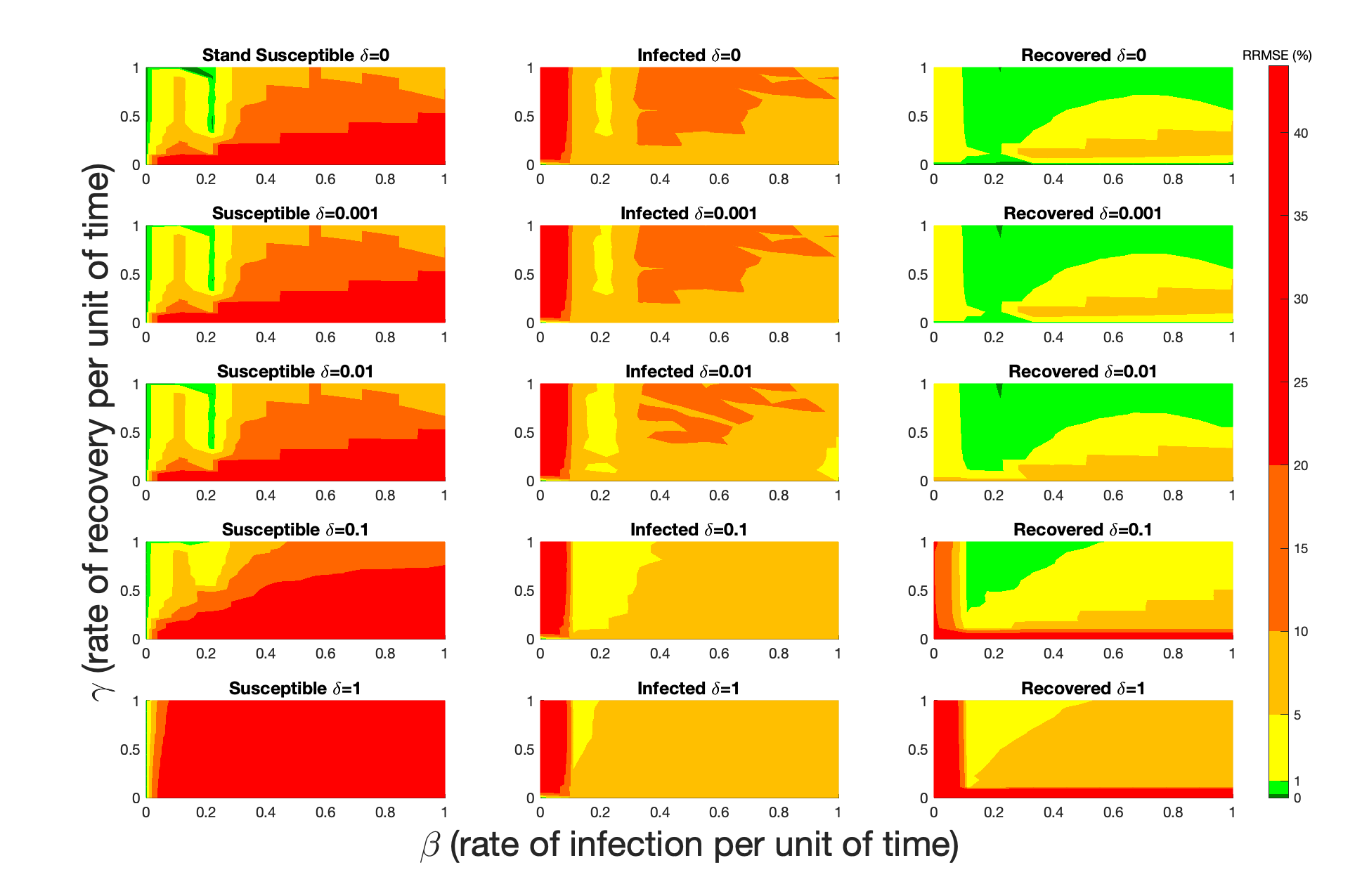}
    \caption{RRMSE percentage across parameter ranges of $[0,1]$ for each respective parameter with $\gamma$ the y-axis of each subfigure, $\beta$ the x-axis of each subfigure, and $\delta$ set at a different fixed value for each subfigure. PN Time Step Per Unit of Time parameter $\tau=1$. Note that red is RRMSE $\leq44\%$ (43.75\% being the max observed RRMSE across all simulations), dark orange is RRMSE $\leq20\%$, light orange is RRMSE $\leq10\%$, yellow is RRMSE $\leq5\%$, light green is RRMSE $\leq1\%$, and dark green is RRMSE $\leq.1\%$}
    \label{fig:Rounding_S}
\end{figure}

\subsection{Spike Petri Nets simulations}
\begin{figure}[H]
    \centering
    \textbf{Spike Layout 1 From Figure \ref{fig:SIRS_PN}}\par
    \includegraphics[width=1\textwidth]{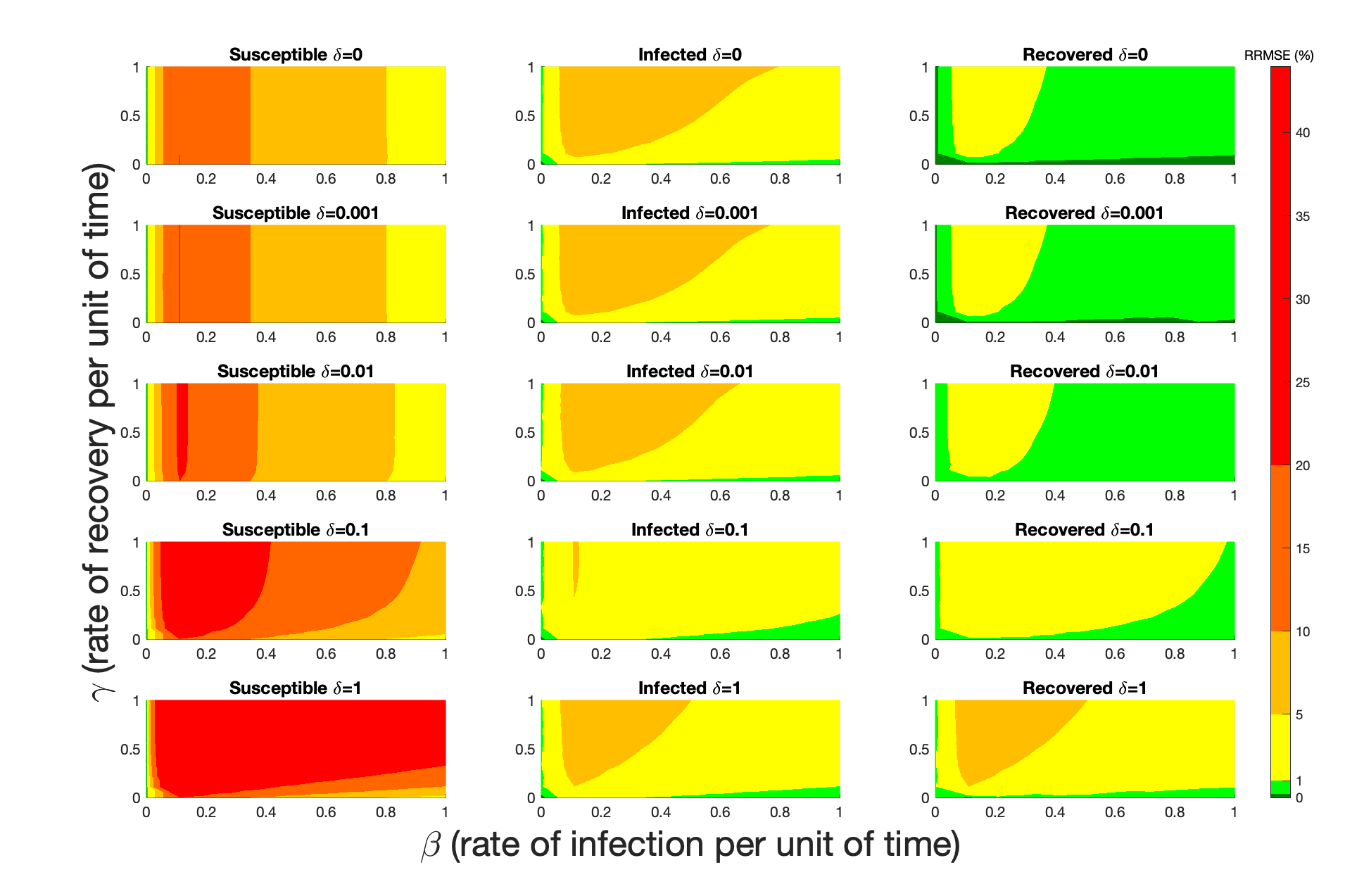}
    \caption{RRMSE percentage across parameter ranges of $[0,1]$ for each respective parameter with $\gamma$ the y-axis of each subfigure, $\beta$ the x-axis of each subfigure, and $\delta$ set at a different fixed value for each subfigure. PN Time Step Per Unit of Time parameter $\tau=1$. Note that red is RRMSE $\leq44\%$ (43.75\% being the max observed RRMSE across all simulations), dark orange is RRMSE $\leq20\%$, light orange is RRMSE $\leq10\%$, yellow is RRMSE $\leq5\%$, light green is RRMSE $\leq1\%$, and dark green is RRMSE $\leq.1\%$}
    \label{fig:Spike_Layout1}
\end{figure}

\begin{figure}[H]
    \centering
    \textbf{Spike Layout 2 From Figure \ref{fig:SIR_v2}}\par
    \includegraphics[width=1\textwidth]{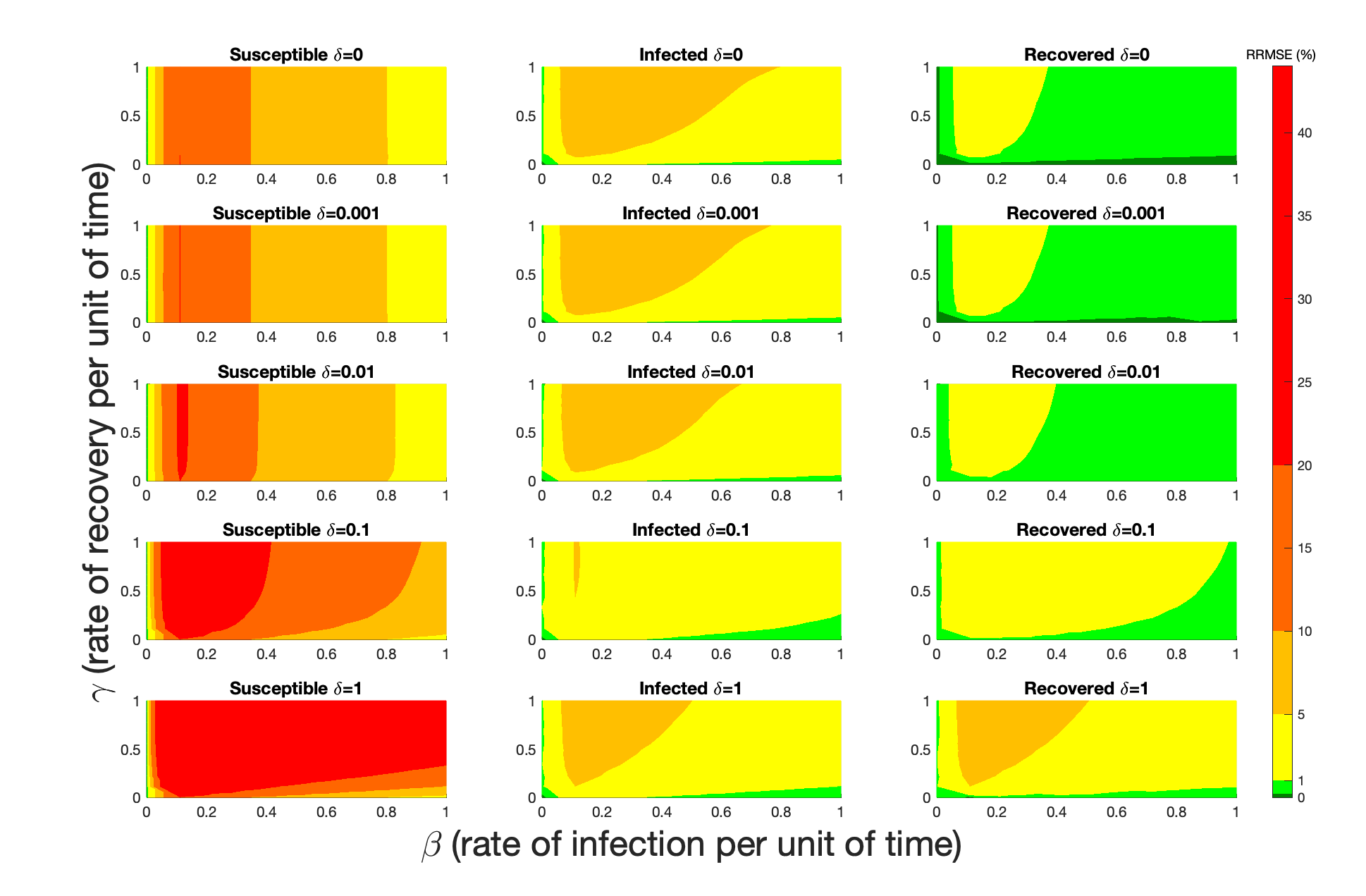}
    \caption{RRMSE percentage across parameter ranges of $[0,1]$ for each respective parameter with $\gamma$ the y-axis of each subfigure, $\beta$ the x-axis of each subfigure, and $\delta$ set at a different fixed value for each subfigure. PN Time Step Per Unit of Time parameter $\tau=1$. Note that red is RRMSE $\leq44\%$ (43.75\% being the max observed RRMSE across all simulations), dark orange is RRMSE $\leq20\%$, light orange is RRMSE $\leq10\%$, yellow is RRMSE $\leq5\%$, light green is RRMSE $\leq1\%$, and dark green is RRMSE $\leq.1\%$}
    \label{fig:Spike_Layout2}
\end{figure}

\subsection{Pop Scalar/$\tau$ runs GPenSIM}
\begin{figure}[H]
    \centering
    \textbf{Population Scalar=4, $\tau=20$}\par
    \includegraphics[width=1\textwidth]{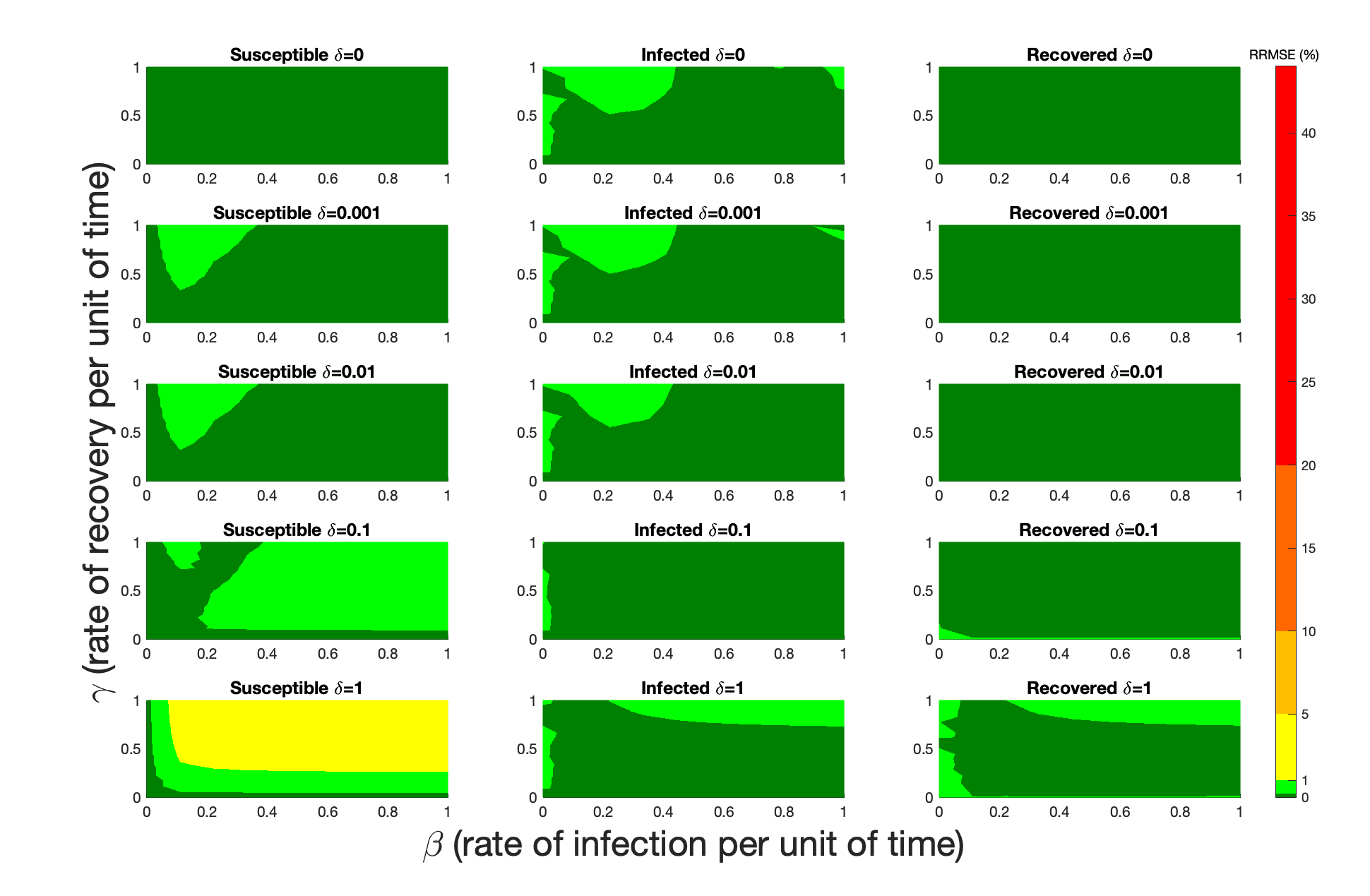}
    \caption{RRMSE percentage across parameter ranges of $[0,1]$ for each respective parameter with $\gamma$ the y-axis of each subfigure, $\beta$ the x-axis of each subfigure, and $\delta$ set at a different fixed value for each subfigure. PN Time Step Per Unit of Time parameter $\tau=1$. Note that red is RRMSE $\leq44\%$ (43.75\% being the max observed RRMSE across all simulations), dark orange is RRMSE $\leq20\%$, light orange is RRMSE $\leq10\%$, yellow is RRMSE $\leq5\%$, light green is RRMSE $\leq1\%$, and dark green is RRMSE $\leq.1\%$}
    \label{fig:PopScalar_4_tau_20_ODE_Comparison}
\end{figure}

\subsection{Spike Stochastic Petri Nets simulations}
\begin{figure}[H]
    \centering
    \textbf{Spike Direct Stochastic PN, Population Scalar=1000}\par
    \includegraphics[width=1\textwidth]{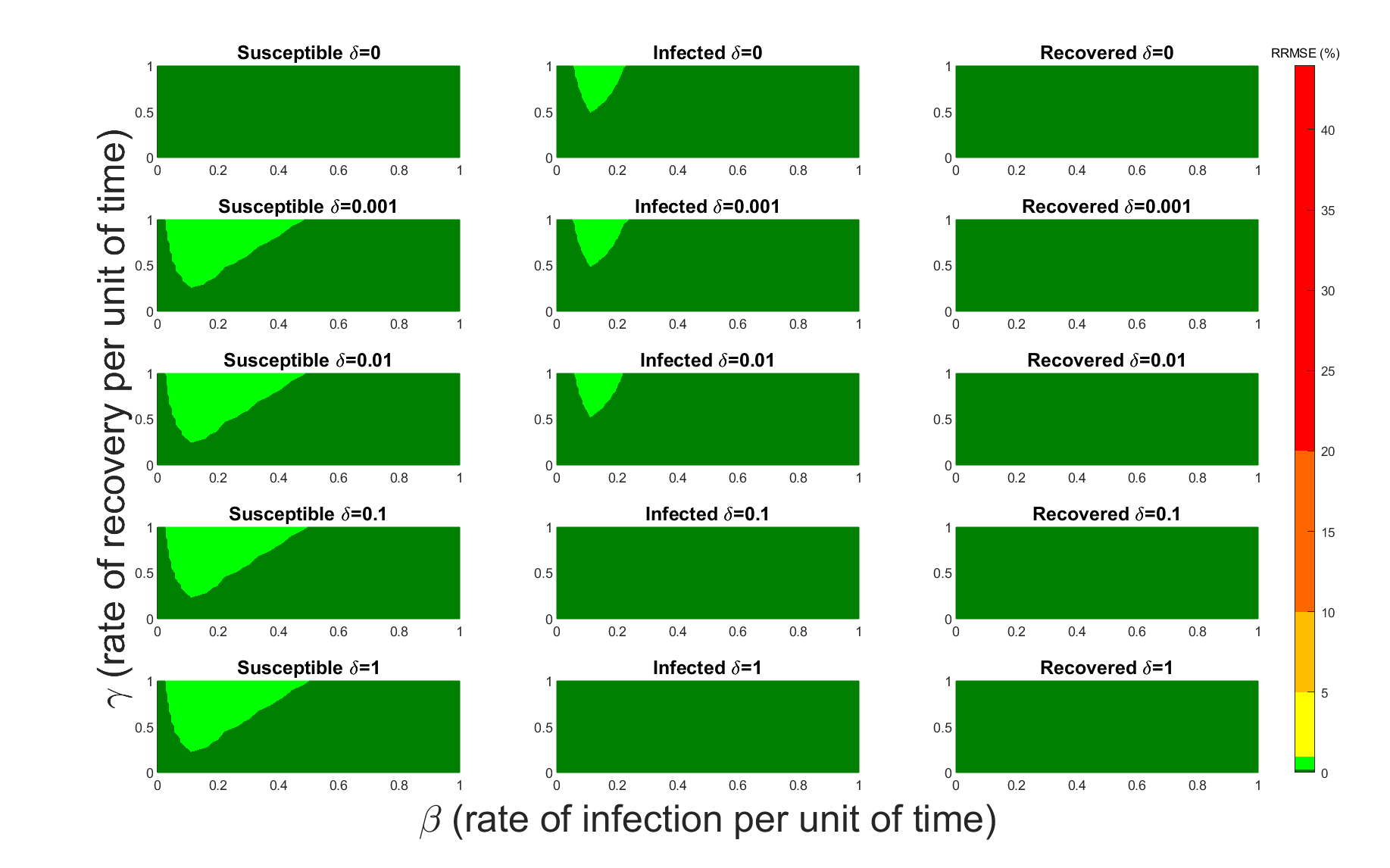}
    \caption{RRMSE percentage across parameter ranges of $[0,1]$ for each respective parameter with $\gamma$ the y-axis of each subfigure, $\beta$ the x-axis of each subfigure, and $\delta$ set at a different fixed value for each subfigure. Note that light orange is RRMSE $\leq10\%$, yellow is RRMSE $\leq5\%$, light green is RRMSE $\leq1\%$, and dark green is RRMSE $\leq.1\%$}
    \label{fig:directpop1000}
\end{figure}

\begin{figure}[H]
    \centering
    \textbf{Spike Tau Leaping Direct Stochastic PN, Population Scalar=1000}\par
    \includegraphics[width=1\textwidth]{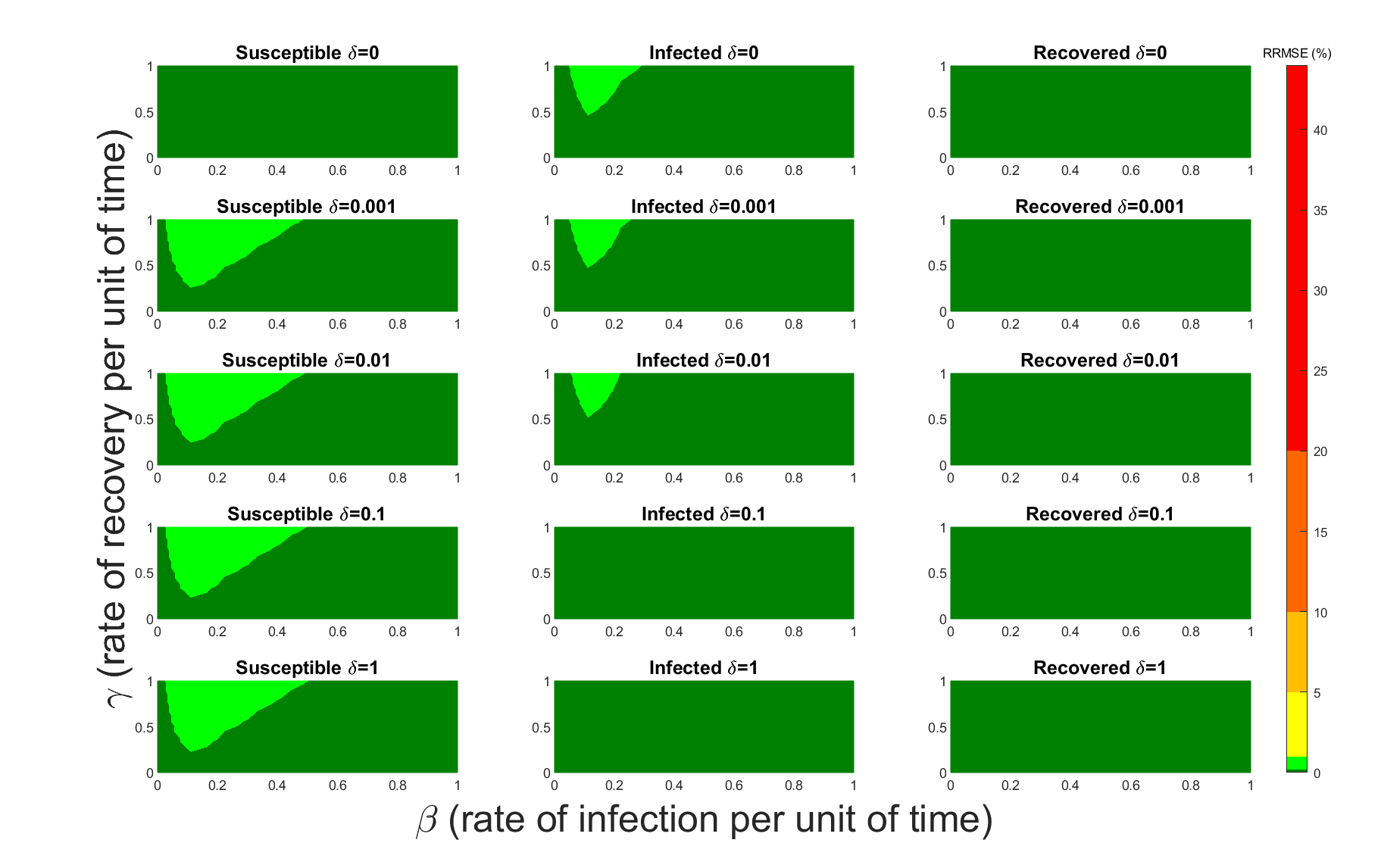}
    \caption{RRMSE percentage across parameter ranges of $[0,1]$ for each respective parameter with $\gamma$ the y-axis of each subfigure, $\beta$ the x-axis of each subfigure, and $\delta$ set at a different fixed value for each subfigure. Note that light orange is RRMSE $\leq10\%$, yellow is RRMSE $\leq5\%$, light green is RRMSE $\leq1\%$, and dark green is RRMSE $\leq.1\%$}
    \label{fig:tauleaping_pop1000}
\end{figure}

\end{document}